\documentclass[preprint]{aastex} 
\usepackage{amsmath,amssymb}

\newcommand{\chandra}{\textit{Chandra}}
\newcommand{\xmm}{XMM \textit{Newton}}
\newcommand{\zpup}{$\zeta$ Pup}
\newcommand{\tori}{$\theta^1$ Ori C}
\newcommand{\vinf}{\ensuremath{v_{\infty}}}
\newcommand{\msunyr}{\ensuremath{\mathrm{M}_\sun~\mathrm{yr}^{-1}}}
\newcommand{\mdot}{\dot{M}}

\begin{document}

\shorttitle{\chandra\ Spectroscopy of \tori} \shortauthors{Gagn\'{e},
Oksala, Cohen, Tonnesen, ud-Doula, Owocki, Townsend, \& MacFarlane}

\title{\chandra\ HETGS Multi-phase Spectroscopy of the Young
Magnetic O Star $\theta^1$ Orionis C}

\author{Marc Gagn\'{e}\altaffilmark{1}, Mary E. Oksala\altaffilmark{1,5},
David H. Cohen\altaffilmark{2}, Stephanie K. Tonnesen\altaffilmark{2,6},
Asif ud-Doula\altaffilmark{3}, Stanley P. Owocki\altaffilmark{4}, 
Richard H.D. Townsend\altaffilmark{4} \&
Joseph J. MacFarlane\altaffilmark{7}}

\email{mgagne@wcupa.edu}

\altaffiltext{1}{West Chester University, Department of Geology and
Astronomy, West Chester PA 19383}

\altaffiltext{2}{Swarthmore College, Department of Physics and
Astronomy, 500 College Ave., Swarthmore PA 19081}

\altaffiltext{3}{North Carolina State University, Department of
Physics, 2700 Stinson Dr., Raleigh NC 27695 }

\altaffiltext{4}{University of Delaware, Bartol Research Institute, 
217 Sharp Laboratory, Newark DE 19716}

\altaffiltext{5}{University of Delaware, Department of Physics \& Astronomy, 
223 Sharp Laboratory, Newark DE 19716}

\altaffiltext{6}{Columbia University, Department of Astronomy, 
1328 Pupin Physics Laboratories, New York NY 10027}

\altaffiltext{7}{Prism Computational Sciences, 455 Science Dr.,
Madison WI 53711}

\begin{abstract}

We report on four \chandra\ grating observations of the oblique
magnetic rotator \tori\ (O5.5~V) covering a wide range of viewing angles with
respect to the star's 1060~G dipole magnetic field.  We employ line-width
and centroid analyses to study the dynamics of the X-ray
emitting plasma in the circumstellar environment, as well as
line-ratio diagnostics to constrain the spatial location, and global
spectral modeling to constrain the temperature distribution and abundances
of the very hot plasma.  We investigate these
diagnostics as a function of viewing angle and analyze them in
conjunction with new MHD simulations of the magnetically channeled
wind shock mechanism on \tori. This model fits all the data
surprisingly well, predicting the temperature, luminosity, and
occultation of the X-ray emitting plasma with rotation phase.
\end{abstract}


\section{Introduction}

\tori, the brightest star in the Trapezium and the primary source of ionization
of the Orion Nebula, is the only O star with a measured magnetic field 
\citep{Donati2002}.  This hot star is an unusually strong and
hard X-ray source \citep{Yamauchi1993,Yamauchi1996,Schulz2000},
with its X-ray flux modulated on the star's 15-day rotation period
\citep{Gagne1997}. Hard, time-variable X-ray emission
is not expected from O stars because they do not possess an outer convective
envelope to drive a solar-type magnetic dynamo.

Rather, \citet{Donati2002}'s detection of a variable longitudinal magnetic
field confirms the basic picture of an oblique magnetic rotator with
a dipolar surface field strength, $B \approx 1060$~G.
As such, \tori, with its relatively strong ($\mdot \approx 4 \times
10^{-7}$ \msunyr) radiation-driven wind, may be the prototype of a new
class of stellar X-ray source: a young, hot star in which a
hybrid magnetic-wind mechanism generates strong and hard, modulated
X-ray emission.

Building on the work of \citet{SB1990} for Ap/Bp stars,
\citet{BM1997a,BM1997b} first
described quantitatively the mechanism whereby a radiation-driven wind
in the presence of a strong, large-scale dipole magnetic field can be
channeled along the magnetic field lines. In this Magnetically
Channeled\footnote{Originally referred to as the magnetically {\em
confined} wind shock model by \citet{BM1997a}, we prefer Magnetically
{\em Channeled} Wind Shock model to emphasize the lack of complete
magnetic confinement of the post-shock wind plasma \citep{UO2002}.}
Wind Shock (MCWS) model, flows from the two hemispheres meet
at the magnetic equator, potentially at high velocity.  This head-on
collision between fast wind streams can lead to very high ($T \gg 10^7$ K)
shock temperatures, much higher than those seen in
observations of normal O stars \citep[e.g., \zpup;][]{Hillier1993}.
X-ray production in O-type supergiants is attributed to the line-force
instability wind shock model \citep{OCR1988,FPP1997},
in which shocks form from interactions between wind streams that are flowing
radially away from the star at moderately different speeds.

What makes the MCWS shocks more efficient than instability-driven wind shocks
is the velocity contrast experienced in the shocks. 
The magnetically-channeled wind streams collide with the cool, dense,
nearly stationary, post-shock plasma at the magnetic equator. 
The rapid deceleration of the wind plasma, and the consequent conversion
of kinetic energy into heat, causes the strong, hard shocks seen
in the MCWS model.
This leads to high levels of X-ray emission since
a large fraction of the wind kinetic energy is converted into shock heating.
In the absence of a strong magnetic field, O-star winds do not produce a
fast wind running into stationary plasma; the shocks correspondingly lead
to lower shock temperatures and lower luminosity.

This picture of a magnetosphere filled with plasma and corotating with the star,
is similar to the picture that has been applied to magnetic Ap/Bp stars
\citep{SB1990,Shore1990,SG2001}. Indeed Babel \& Montmerle's first
paper on the subject was an application to the Ap star IQ Aur
{\citep{BM1997a}. However, a major difference between the model's
application to Ap/Bp stars and to young O stars is that the winds of O
stars are significantly stronger, representing a non-negligible
fraction of the star's energy and momentum output. This hybrid
magnetic wind model thus is more than a means of explaining the
geometry of circumstellar material, but also can potentially explain
the intense hard X-ray emission from young, magnetized hot stars (three
orders of magnitude greater for \tori\ than for IQ Aur). As has
been pointed out in the context of the magnetic Ap/Bp stars, if the
dipole field is tilted with respect to the rotation axis, then the
rotational modulation of our viewing angle of the magnetic pole and
its effect on observed flux levels and line profiles provides a
important diagnostic tool. We exploit this fact by making
several \chandra\ observations of \tori\ at different rotational
phases, in order to study the dynamics and geometry of the hot, X-ray
emitting plasma in the circumstellar environment of this young,
magnetized O star and thus to assess the mechanism responsible for
generating the unusual levels of X-ray emission in \tori.

The time independent nature of the original Babel \& Montmerle model
leads to a steady-state configuration in which a post-shock
``cooling disc'' forms at the magnetic equator. 
In their model, this disk is essentially an
infinite mass sink.  This has an effect on the assumed
X-ray absorption properties of the disk and it also
affects the steady state spatial, temperature, and velocity
distribution of the hot, X-ray emitting plasma, thus affecting
the quantitative predictions of observables in unphysical ways. 
\citet{BM1997b} further assume a rigid magnetic geometry within
the magnetosphere, thereby imposing a
pre-specified trajectory to the wind streams; namely they are forced
to flow everywhere along the local dipole field direction.  In
reality, if the wind kinetic energy is large enough, the magnetic
field should be distorted.  In the extreme case of a field that is
(locally) weak compared to the wind kinetic energy, the initially
closed field lines will get ripped open and the flow (and field lines)
will be radial, and the focusing of the wind flow toward the equator
and the associated shock heating will be minimized.  
\citet{BM1997b} recognized these limitations and speculated about
the effects of relaxing the strong-field and steady-state assumptions
\citep[see also][]{Donati2002}.

To explore these effects, \citet{UO2002} performed magneto-hydrodynamic
(MHD) simulations of radiation-driven winds in the presence of a magnetic field.
Their results confirmed the basic paradigm of the Babel \& Montemerle model 
wherein a magnetic field channels the wind material towards the magnetic
equator where it shocks and cools. If the field is strong enough it
can even confine the wind. The overall degree of such a confinement can
be determined by a single dimensionless parameter, $\eta_{\star}$,
which represents the ratio of the magnetic energy density to the wind
kinetic energy density.
\begin{equation}
\eta_{\star} = \frac{B^2_{\rm eq} R^2_{\star}}{\mdot \vinf},
\label{eta}
\end{equation}
where $B_{\rm eq}$ is the equatorial dipole magnetic intensity at the
stellar surface, $R_{\star}$ is the stellar radius, $\mdot$ is the 
mass-loss rate, and $\vinf$ is the terminal wind speed.
For $\eta_{\star} \lesssim 1$, the field is relatively weak
and the wind dominates over the magnetic field by ripping the
field lines into a nearly radial configuration. Nonetheless, the field
still manages to divert some material towards the magnetic equator leading to a
noticeable density enhancement. Because the magnetic field energy density falls
off significantly faster than the wind kinetic energy density, at large radii,
the wind always dominates over the field, even for $\eta_{\star} \gg 1$.
The MHD simulations reveal some phenomena not predicted by the static
Babel \& Montmerle model. Above the closed magnetosphere near the magnetic
equator where the field lines start to open up, the magnetically channeled
wind still has a large latitudinal velocity component, leading to strong shocks.

The simulations show shock-heated wind plasma above and below the magnetic
equatorial plane where it cools and then falls along closed magnetic
field lines. This cool, dense, post-shock plasma forms an unstable
disk. Mass in the disk builds up until it can no longer be supported by
magnetic tension, at which time it falls back down onto the photosphere. 
The star's rotation and the obliquity of the magnetic axis
to the rotation axis lead to periodic occultation of some of the 
X-ray emitting plasma. All of these phenomena should lead to observable,
viewing-angle dependent diagnostics: temperature distribution,
Doppler line broadening, line shifts, and absorption.

The initial 2D MHD simulations performed by \citet{UO2002} were
isothermal, and the shock strength and associated heating were
estimated based on the shock-jump velocities seen in the MHD output.
For the analysis of the \chandra\ HETGS spectra discussed in
this paper, however, new simulations were performed that include adiabatic
cooling and radiative cooling as well as shock heating in the MHD
code's energy equation.  In this way, we make detailed,
quantitative predictions of plasma temperature and hence X-ray
emission for comparison with the data. Furthermore, the changes in gas
pressure associated with the inclusion of heating and cooling 
has the potential to affect the dynamics and thus the shock physics and
geometry of the magnetosphere. We thus present the
most physically realistic modeling to date of X-ray production in the
context of the MCWS model.

In this paper we present two new observations of \tori\ taken with the
HETGS aboard \chandra, which we analyze in conjunction with two GTO
observations, previously published by \citet{Schulz2000,Schulz2003}.
These four observations provide good coverage of the rotational phase
and hence sample a broad range of viewing angles with respect to the
magnetosphere. We apply several spectral diagnostics, including
line-width analysis for plasma dynamics, line-ratio analysis for
distance from the photosphere, differential emission measure for the
plasma temperature distribution, and overall X-ray spectral
variability for the degree of occultation and absorption, to these
four data sets.  In this way, we characterize the plasma properties
and their variation with phase, or viewing angle, allowing us to
compare these results to the output of MHD simulations and assess the
applicability of the MCWS model to this star.

In \S2 we describe the data and our reduction and analysis procedure.
In \S3 we discuss the properties of \tori.
\S4 we describe the MHD modeling and associated diagnostic synthesis.
In \S5 we report on the
results of spectral diagnostics and their variation with phase.
In \S6 we discuss the points of agreement and disagreement between the
modeling and the data, critically assessing the spatial distribution
of the shock-heated plasma and its physical properties in light of the
observational constraints and the modeling.  We conclude with an
assessment of the applicability of hybrid magnetic-wind models to
\tori\ in \S7.

\section{\chandra\ Observations and Data Analysis}

The \chandra\ X-ray Observatory observed the Trapezium with the
High Energy Transmission Grating (HETG) spectrometer twice during
Cycle 1 and twice during Cycle 3. The Orion Nebula Cluster was also
observed repeatedly during Cycle 4 with the ACIS-I camera with no gratings.
Table~1 lists each observation's Sequence Number, Observation ID (OBSID),
UT Start and End Date, exposure time (in kiloseconds), average phase, and
average viewing angle (in degrees). The phase was calculated using a period 
of 15.422 days and a zero-point MJD=48832.5 \citep{Stahl1996}.
Assuming a centered oblique magnetic rotator geometry (see Figure~\ref{f1}),
the viewing angle $\alpha$ between the line of sight and the magnetic pole
is specified by:
\begin{equation}
\cos \alpha = \sin \beta \cos \phi \sin i + \cos \beta \cos i,
\label{vangle}
\end{equation}
where $\beta=42\arcdeg$ is the obliquity (the angle between the rotational
and magnetic axes), $i=45\arcdeg$ is the inclination angle, and $\phi$
is the rotational phase of the observation (in degrees).
In Fig.~\ref{f1}, the arrows at each viewing angle point to Earth.
As the star rotates, the viewing angle varies from
$\alpha\approx 3\arcdeg$ (nearly pole-on) at phase $\phi=0.0$ to
$\alpha\approx 87\arcdeg$ (nearly edge-on) at phase $\phi=0.5$.
The four grating observations were obtained
at $\alpha\approx 4\arcdeg, 40\arcdeg, 80\arcdeg$, and $87\arcdeg$,
thus spanning the full range of possible viewing angles.
An online animation\footnote{http://www.bartol.udel.edu/t1oc}
illustrates our view of the magnetosphere as a function of rotational phase.

The \chandra\ HETG data were taken with the ACIS-S camera in TE-mode.
During OBSID 3 and 4, ACIS CCDs S1-S5 were used; S0 was not active.
During OBSID 2567 and 2568, ACIS CCDs S0-S5 were used.
The data were reduced using standard threads in CIAO version 2.2.1
using CALDB version 2.26.
Events with standard ASCA grades 0, 2, 3, 4, and 6 were retained.
The data for CCD S4 were destreaked because of a flaw in the serial readout
which causes excess charge to be deposited along some rows. The data
extraction resulted in four first-order spectra for each observation:
positive and negative first-order spectra for both the medium-energy
grating (MEG) and high-energy grating (HEG).

A grating Auxiliary Response File (ARF) was calculated for each
grating spectrum. The negative and positive first order
spectra were then coadded for each observation using the CIAO program
ADD\_GRATING\_ORDERS. The four sets of MEG and HEG spectra are shown in
the top four panels of Figures \ref{f2} and \ref{f3}. The first-order MEG
spectra from each of the four observations were then co-added to create
the spectrum in the bottom panel of Fig.~\ref{f2}.
The co-added HEG spectrum is shown in the bottom panel of Fig.~\ref{f3}.
The MEG gratings have lower dispersion than the HEG gratings.
However, the MEG is more efficient than the HEG above 3.5~\AA.
The added spectral resolution of the HEG is useful for resolving line
profiles and resolving lines that may be blended in the MEG spectrum.
Because the HEG and MEG wavelength solutions were in excellent agreement,
the MEG and HEG spectra were fit together to increase signal-to-noise.

The spectra show very strong line and continuum emission below
10~\AA\ indicating a very hot, optically thin plasma. The relatively
low line-to-continuum ratio from 9--12~\AA\ suggests sub-solar Fe abundance.
Photons with $\lambda>24$~\AA\ are nearly entirely absorbed
by neutral hydrogen, helium and metals in the wind of \tori, in the neutral
lid of the Orion Nebula and in the line-of-sight ISM.

We analyzed the extracted spectra in several different ways.  In \S5
we discuss the diagnostics that we bring to bear on the determination
of the properties of the hot plasma, but first we describe our data
analysis procedures.
Our first goal was to determine
centroid wavelengths, intensities and non-instrumental line widths for the
eighty or so brightest emission lines.
Our second goal was to determine the basic physical
properties of the X-ray emitting plasma.

\subsection{Global Spectral Fits}

In order to determine the temperatures, emission measures, and abundances
of the emitting plasma and the column density of the overlying gas, we fit
the HEG/MEG spectra using the
ISIS\footnote{ISIS: Interactive Spectral Interpretation System, version 1.1.3}
software package. We obtained acceptable fits to the 1.8--23~\AA\ HEG and MEG
spectra using a two-temperature variable-abundance
APED\footnote{The Astrophysical Plasma Emission
Database is an atomic data collection to calculate plasma continuum +
emission-line spectra. See \citet{Smith2001} and references therein.}
emission model with photo-electric absorption from neutral ISM gas 
\citep{MM1983}. The free parameters
were the column density, the two temperatures, the two normalizations,
the redshift, the turbulent velocity, and the abundances of
O, Ne, Mg, Si, S, Ar, Ca and Fe.

We note that a number of emission-measure profiles EM(T) can be
used to fit the spectra, all giving rise to similar results: a dominant
temperature component near 30~MK, and a second temperature peak near 8~MK.
We chose the simplest model with the fewest free parameters that adequately
fit the data: a two-temperature, variable abundance APEC model in ISIS.
We also note that with all the thermal emission models we tried,
non-solar abundances were needed to correctly fit the emission lines.

As we shall discuss in \S5, we find small, but significant, shifts in the line
centroids. The ISIS plasma emission model has one advantage compared to those in
SHERPA and XSPEC: it explicitly and consistently accounts for thermal,
non-thermal, and instrumental line broadening.
The width of a line corrected for instrumental broadening
is the Doppler velocity, $v_D$, as defined by \citet{RL1979}:
\begin{equation}
v_D = \frac{c}{\lambda_0}{\Delta\lambda_D} = \frac{c}{\lambda_0} \frac{{\rm FWHM}}{\sqrt{\ln 16}}
= { \left( \frac{2kT}{m} + \xi^2 \right) } ^ {\frac{1}{2}},
\label{vd}
\end{equation}
where $v_{\rm th} = \sqrt{\frac{2kT}{m}}$ is the thermal broadening component
and $\xi$ is the rms of the turbulent velocities.
$\xi$ is used simply as a standard means of parameterizing
the component of the line width caused by non-thermal bulk motion
(assumed to be Gaussian).
As we discuss further in \S5, the non-thermal, non-instrumental width
parametrized by $\xi$ is probably the result of
bulk motion of shock-heated wind plasma, and not, strictly speaking ,
physical microturbulence as described by \citet{RL1979}.
Note that $v_D$ and $\xi$ thus defined are
related to $v_{\rm HWHM}$, often used as a velocity-width
parameter when broadening is due to a stellar wind, as
$v_{\rm HWHM} = 0.83v_D$.

The spectral anaylsis showed no obvious changes in abundance with phase.
The best-fit abundances
and $1\sigma$ uncertainties are shown in the ``Global Fit'' column of Table 5.
These abundances were then fixed for the fits at each rotational phase.
Those best-fit parameters and $1\sigma$ uncertainties are shown in Table 6.
The most notable result of the global fitting is that most of the plasma
is at $T\approx30$~MK and that the temperatures do not vary significantly with
phase. The phase-to-phase spectral variations seen in Figs.~\ref{f2} and
\ref{f3} appear to be caused by changes in visible emission measure.
One disadvantage of this type of global fitting is that the
thermal emission models currently in ISIS, SHERPA and XSPEC do not account
for the effects of high density and/or UV radiation. For this, we model
the individual line complexes (see \S5.2).

\subsection{Line Profile Analysis}

SHERPA, the CIAO spectral fitting program, was used to fit the
24 strongest line complexes of O, Ne, Mg, Si,
S, Ar, Ca, and Fe in the 1.75 to 23 \AA\ range.
In the 3--13 \AA\ region, the HEG and MEG spectra were
simultaneously fit for each complex. Below 3 \AA\ the MEG spectra have low S/N
and only the HEG spectra were used. Above 13 \AA\ the HEG
spectra have low S/N and only the MEG spectra were used.
Because each line complex
typically contains one to eight relatively bright lines,
multiple Gaussian components were used to model the brightest lines
in a complex (as estimated by their relative
intensities at 30~MK in APED).
Emission from continua and very faint lines across each complex was modeled
simply as an additive constant.
We note, e.g., that unresolved doublets were modeled as two Gaussians, with
the emissivity of the fainter of the pair tied to the emissivity of the
brighter. To further minimize the number of free parameters, the centroid
wavelengths and widths of all lines in a complex were tied to that of the
brightest line in that complex.

For example, a typical He-like line complex
like \ion{Si}{13} (see Figure~\ref{f12}c) was modeled with six Gaussians:
the \ion{Si}{13} resonant (singlet), intercombination (doublet),
and forbidden (singlet) lines plus the \ion{Mg}{12} resonant doublet.
This particular complex had six free parameters: one constant, one line width,
one line shift (relative to APED) and three intensities (\ion{Si}{13} resonance,
one \ion{Si}{13} intercombination, and \ion{Si}{13} forbidden).
In this case, the \ion{Mg}{12} emission was blended with the \ion{Si}{13}
forbidden line. In these cases, blending leads to higher uncertainties
in the line parameters.
In this example, the \ion{Mg}{12} line was accounted for by tying its
flux to the flux of the \ion{Si}{13} resonance line, scaled by the emissivity
ratio of those lines in APED and the Mg/Si ratio from globally fitting
the HEG/MEG spectra.

The resulting SHERPA fits are shown in
the first nine columns of Table~2. The ion, transition, and rest wavelength
$\lambda_0$ are from APED. Free parameters have associated $1\sigma$ errors,
tied parameters do not.
As expected, line and continuum fluxes were approximately 30\% higher
during OBSID 2567 (viewing angle $\alpha=4\arcdeg$) than during OBSID 2568
($\alpha=87\arcdeg$), with intermediate fluxes during OBSID 3 and 4.

Because SHERPA does not explicitly account for thermal and non-thermal line
broadening, the individual line complexes were fit a second time with ISIS to
derive the rms turbulent velocities and abundances. $\xi$ and the relevant
abundance parameters for a given line complex were allowed to vary while the
emission and absorption parameters (temperatures, normalizations,
other abundances, and column density) were frozen.
The resulting abundances and $1\sigma$ uncertainties are listed in the
``Line Fit'' column of Table 5. The Fe abundance was determined from
numerous ionization stages in six line complexes. For Fe, $\sigma$ is the
standard deviation of these six Fe abundance fits.
The individual rms turbulent velocities $\xi$ are listed in the last two
columns of Table 2.

Summarizing the results of the spectral fitting, we conclude that some
elements (especially Fe) have non-solar abundances, most of the plasma
is at a temperature around 30~MK, the lines are resolved but much narrower
than the terminal wind velocity, and only the overall emission measure
appears to vary significantly with phase.

We note that the zeroth-order ACIS-S point-spread function of \tori\ is
substantially broader than nearby point sources and substantially
broader than the nominal ACIS PSF.  We have taken the best-fit model from
ISIS and made MARX simulations of the zeroth-order PSF. The simulations show
that the excess broadening in the zeroth-order image is a result of
pileup in the ACIS-S CCD. Pixels in the core of the PSF register
multiple events occurring within a single 3.2~s readout as a single
higher-energy count. In the wings of the PSF, the count rate per pixel
is much lower and pileup is reduced. Hence the spatial profile is
less peaked than it would be without pileup. Because the count rate
per pixel in the first-order lines is much lower than in zeroth order,
the lines in the dispersed spectra are not broadened by pileup.
We conclude that the measured line velocities are {\em not} caused by
instrumental effects. Non-intrumental Gaussian line shapes provide
adequate fits to the data. Non-Gaussian line shapes, while perhaps
present, could only be detected with higher S/N data.

\section{Properties of \tori}

In order to put the results of the spectral fitting and the
MHD modeling into a physical context, we must first specify the
stellar and wind parameters of \tori\ as well as elaborate on the
geometry and distribution of circumstellar matter, as constrained by
previous optical and UV measurements.
The spectral type of the star is usually taken to be O6~V or O7~V,
although its spectral variability was first noted by \citet{Conti1972}.
\citet{Donati2002} assume a somewhat hot effective temperature,
$T_{\rm eff} = 45500$ K, which is based on the O5.5~V type used by
\citet{HP1989}, which itself is an average of Walborn's (1981) range of O4-O7.
\citet{Walborn1981} showed that lines often used to determine spectral type
(\ion{He}{1} $\lambda4471$, \ion{He}{2} $\lambda 4542$ and $\lambda 4686$)
are highly variable on time scales of about a week, suggesting a
period ``on the order of a few weeks''. He compared \tori's
line-profile variations to those of the magnetic B2~Vp star
$\sigma$~Orionis~E,
even noting then that \tori's X-ray emission, recently discovered
with the {\it Einstein} Observatory,
may be related to its \ion{He}{2} $\lambda4686$ emission.

Based on a series of H$\alpha$ and \ion{He}{2} $\lambda 4686$ spectra
obtained over many years, \citet{Stahl1993,Stahl1996} and \citet{Reiners2000}
have determined that the inclination angle of \tori\ is $i \approx 45\arcdeg$
and the rotational period $P = 15.422\pm 0.002$ days.
The inclination angle estimate is based on arguments about
the symmetry of the light curve and spectral variability as well as
plausibility arguments based on the inferred stellar radius.  
As expected from its long rotation period, \tori's projected rotational velocity
is quite low, with \citet{Reiners2000} finding a value of
$v\sin i = 32 \pm 5$ km s$^{-1}$ using the \ion{O}{3} $\lambda5592$ line core. 

The Zeeman signature measurements of \citet{Donati2002}
indicate a strong dipole field, with no strong evidence for higher
order field components.
The five sets of Stokes parameters indicate a magnetic obliquity 
$\beta \approx 42\arcdeg$ if $i = 45\arcdeg$. These numbers then imply a
photospheric polar magnetic field strength of $B_{\mathrm p} = 1060 \pm 90$~G.
The obliquity and polar field strength take
on modestly different values if the inclination angle is assumed to be
as small as $i=25\arcdeg$ or as large as $i=65\arcdeg$.

As a result of these various determinations of stellar properties,
we adopt two sets of stellar parameters that likely
bracket those of \tori.  These are listed in Table 3,
and are based on the PHOENIX spherical NLTE model atmospheres of
\citet{Aufdenberg2001}.

The theoretical wind values listed in Table 3 are
consistent with the observed values of $\mdot = 4 \times 10^{-7}$
\msunyr\ \citep{HP1989} and $\vinf \approx 2500$ km s$^{-1}$
\citep{Stahl1996}. The stellar wind properties, as seen in UV resonance
lines, are modulated on the rotation period of the star
\citep{WN1994,Reiners2000}, which, when combined with variable H$\alpha$
\citep{Stahl1996} and the new measurement of the variable magnetic field
\citep{Donati2002}, lead to a coherent picture of an oblique magnetic
rotator, with circumstellar material channeled along the magnetic equator.
This picture is summarized in Fig.~\ref{f1},
which also indicates the viewing angle relative to
the magnetic field for our four \chandra\ observations.

When we see the system in the magnetic
equatorial plane, the excess UV wind absorption is greatest.
When we see the system magnetic pole on, the UV excess is at its minimum,
while the longitudinal magnetic field strength is maximum, 
as are the X-ray and H$\alpha$ emission \citep{Stahl1993,Gagne1997}.  
These results are summarized in Figure~\ref{f4}, where we show fourteen
days of data from the 850-ks \chandra\ Orion Ultra-Deep Project.
These data were obtained with the ACIS-I camera with no gratings.
Because \tori's count rate far exceeds the readout rate of the ACIS-I CCD,
the central pixels of the source suffer from severe pileup.
For Fig.~\ref{f4}b, counts were extracted from an annular region in
the wings of the PSF where pileup was negligeable
As a result, the \chandra\ light curve does not show
the absolute count rate in the ACIS-I instrument.
The light curve shows long-term variability induced by rotation and occultation
and statistically significant short-term variability that may reflect the
dynamic nature of the magnetically channeled wind shocks.

These four different observational results are consistent with this picture
provided that the H$\alpha$ emission is produced in the post-shock
plasma near the magnetic equatorial plane.
\citet{SF2005} have re-analyzed the 22 high-dispersion IUE
spectra of \tori\ and re-interpreted the time-variable UV and optical
line profiles in terms of the magnetic geometry of \citet{Donati2002}.
\citet{SF2005} find that the \ion{C}{4} and \ion{N}{5} lines at phase
0.5 (nearly edge-on) are characterized by high, negative-velocity
absorptions and lower, positive-velocity emission, superposed on the
baseline P~Cygni profile. Similarly, some of the optical Balmer-line
and \ion{He}{2} emissions appear to be produced by post-shock gas falling
back onto to the star. In the following
sections, we further constrain and elaborate on this picture by using
several different X-ray diagnostics in conjunction with MHD modeling.

We note finally that \tori\ has at least one companion. \citet{Weigelt1999}
find a companion at 33 mas via infrared speckle interferometry.
\citet{Donati2002} estimate its orbital period to be 8--16~yr and
\citet{Scherlt2003} estimate its spectral type as B0-A0.
This places the companion at 15~AU, too far to significantly interact with
the wind or the magnetic field of \tori.
The amplitude and timescale of the X-ray luminosity variations cannot be
caused by colliding-wind emission with the companion, by
occultation by the companion, or by the companion itself \citep{Gagne1997}.

\section{Magneto-Hydrodynamical Modeling of the Magnetically Channeled Wind of \tori}
\subsection{Basic Formulation}

As discussed in the Introduction, we have carried out MHD simulations
tailored specifically for the interpretation of the \tori\ X-ray data 
presented in this paper.  
Building upon the full MHD models by \citet{UO2002}, 
these simulations are self-consistent in the sense that
the magnetic field geometry, which is initially specified as a pure
dipole with a polar field strength of 1060 G, is allowed to adjust in
response to the dynamical influence of the outflowing wind.  
As detailed below, a key improvement over the isothermal simulations  
of \citet{UO2002} is that we now use an explicit energy equation to follow 
shock heating of material to X-ray emitting temperatures.

The overall wind driving, however, still follows a local CAK/Sobolev 
approach that suppresses the ``line deshadowing instability'' 
for structure on scales near and below the Sobolev length,
$l \approx v_{\rm th} r/v \approx 10^{-2}r$, 
with $r$ the radius and $v$ and $v_{\rm th}$ the flow and thermal speeds
\citep{LS1970, MHR1979, OR1984, OR1985}.
For non-magnetic winds, 1-D simulations using a non-local, non-Sobolev
formulation for the line-force \citep{OCR1988, OP1999, RO2002}
show this instability can lead to embedded weak shock structures that might 
reproduce the soft X-ray emission \citep{Feldmeier1995, FPP1997}
observed in hot stars like $\zeta$~Pup \citep{WC2001,Kramer2003}.
However, it seems unlikely that such small-scale, weak-shock 
structures can explain the much harder X-rays oberved from \tori.
Given, moreover, the much greater computational expense of the non-local 
line-force computation, especially for the 2D models computed here
\citep{DO2003, DO2005}, 
we retain the much simpler, CAK/Sobolev form for the line-driving.
In the absence of a magnetic field, such an approach just relaxes to 
the standard, steady-state, spherically symmetric CAK solution.
As such, the extensive flow structure, and associated hard X-ray 
emission, obtained in the MHD models here are a direct consequence of 
the wind channeling by the assumed magnetic dipole originating from the
stellar surface.

Since the circa 15-day rotation period of \tori\ is much longer than the 
characteristic wind flow time of a fraction of a day, we expect 
that associated centrifugal and coriolis terms should have limited effect on 
either the wind dynamics or magnetic field topology.
Thus, for simplicity, and to retain the computational efficiency of 
axisymmetry, the numerical simulations here formally assume no rotation.
However, in comparing the resulting X-ray emission to observations, 
we do account for the rotational modulation arising from the change in 
observer viewing angle with rotational phase, using the 
estimated inclination angle $i \approx 45\arcdeg$ between the
rotation axis and the observer's line of sight, and obliquity
$\beta \approx 42\arcdeg$.

\subsection{Energy Balance with Shock Heating and Radiative Cooling}

In order to derive quantitative predictions for the X-ray emission 
from the strong wind shocks that arise from the magnetic confinement, 
the simulations here extend the isothermal models of \citet{UO2002} 
to include now a detailed treatment of the wind energy balance, 
accounting for both shock heating and radiative cooling.
Specifically, the total time derivative ($D/Dt$) of wind internal energy 
density $e$ 
(related to the gas pressure $p$ by $e=p/(\gamma -1) = (3/2)p$ 
for a ratio of specific heats $\gamma = 5/3$ appropriate for a 
monatomic gas) is computed from
\begin{equation}
    \rho  \, \frac{D (e/\rho)}{Dt} = - \rho \nabla \cdot p + H - C \, ,
\end{equation}
where $\rho$ is the mass density, and $H$ and $C$ represent 
volumetric heating and cooling terms.
In steady hot-star winds, the flow is kept nearly isothermal at 
roughly the stellar effective temperature (few $\times 10^{4}$~K) through the 
near balance between heating by photoionization from the 
underlying star and cooling by radiative emission
\citep[mostly in lines, see, e.g.,][]{Drew1989}.
In such steady outflows, both the heating and cooling terms are intrinsically 
much larger than the adiabatic cooling from the work expansion, 
as given by the first term on the right-hand-side of the energy equation.

But in variable, structured winds with compressive shocks, this first term 
can now lead to strong compressive heating, with some material now 
reaching much higher, X-ray emitting temperatures.
Through numerical time integrations of the energy equation,
our MHD simulations account 
directly for this compressive heating near shocks, imposing also a floor 
value at the photospheric effective temperature 
to account for the effect of photoionization heating on the relatively 
cool, expanding regions. In the simulations, the temperature of the far wind
is $T=45700$~K, slightly warmer than the O5.5 photosphere (see Table~3).

The shock-heated material is cooled by the radiative loss term, which 
is taken to follow the standard form for an optically thin plasma,
\begin{equation}
C  \,  = \,  n_{\rm e}\,  n_{\rm H} ~ \mathcal{L}(T) \, ,
\label{cool}
\end{equation}
where  $n_{\rm e}$ and $n_{\rm H}$ are the numbers densities for 
electrons and hydrogen atoms, and $T$ is the electron temperature.
 
A proper radiative cooling calculation must include many different ionic 
species with all the possible lines \citep{RCS1976,MB1981}, with the exact 
form of the cooling function $\mathcal{L}(T)$ depending on 
the composition of heavy elements.
Detailed calculations for the typical case of solar abundances 
show that, at low temperatures, the cooling function $\mathcal{L}(T)$ increases
approximately monotonically with temperature, reaching a maximum, 
$\mathcal{L} \approx 3\times 10^{-22}$~erg~cm$^{3}$~s$^{-1}$,
at $T\approx2\times10^5$~K \citep{RCS1976}, where the cooling
is dominated by collisional excitation of abundant lithium-like atoms.
Above $T\approx 2\times10^5$~K, $\mathcal{L}(T)$ decreases with temperature,
except for some prominent bumps at approximately 2 and 8~MK.
Above $T\approx3\times10^7$~K, thermal brehmsstrahlung radiation 
dominates and the cooling function turns over again,
increasing with temperature.
In our simulations, we follow the tabulated form of the \citet{MB1981} 
cooling curve based on this solar-abundance recipe.  

\subsection{Simulations Results for Temperature and Emission Measure}

One of the key results of \citet{UO2002} is that the overall degree 
to which the wind is influenced by the magnetic field depends largely on
a single parameter, the magnetic confinement parameter, $\eta_\star$,
defined by equation~\ref{eta}.
For our MHD simulations here of the magnetic wind confinement in \tori, 
we use $\eta_{\star} = 7.5$, based on an assumed 
stellar radius $R_\star\approx 9R_\sun$ (see Table~3),
$B_{\rm eq}=\frac{1}{2} B_{\mathrm p} \approx 530$~G \citep{Donati2002}.
The simulation produces an average mass-loss rate
${\dot M}\approx 4\times10^{-7} M_{\odot}$~yr$^{-1}$ \citep{HP1989},
and terminal wind speed $v_{\infty}\approx 1400$~km~s$^{-1}$.
Thus, we anticipate that, at least qualitatively, the
simulation results for \tori\ should be similar to the strong magnetic 
confinement case ($\eta_{\star} = 10$) shown in Fig.\ 9 of 
\citet{UO2002}, allowing however for differences from the inclusion now 
of a full energy equation instead of the previous assumption of a
strictly isothermal outflow.

We note that \tori's observed
terminal wind speed derived from IUE high-resolution spectra
\citep{HP1989,Prinja1990,WN1994} varies significantly, with
$v_{\infty}\approx 510$~km~s$^{-1}$ at rotational phase 0.1,
when the wind absorption is weak (see top panel of Fig.~4).
\tori's wind speed and mass-loss rate are lower than those of
other mid-O stars and lower than expected from model atmosphere
calculations (see Table~3).
In the interpretation of \citet{SF2005} and \citet{Donati2002},
most of the wind is channeled to the magnetic equatorial region. Thus,
${\dot M}$ and $v_{\infty}$ are low near X-ray maximum (phase 0.0) when the
star is viewed magnetic pole-on.
It is also worth noting that \citet{Prinja1990} measure a narrow absortion-line
component velocity $v_{\rm NAC} \approx 350$~km~s$^{-1}$ at phase 0.1,
comparable to the Doppler velocity of the X-ray lines discussed in \S5.
In contrast, \citet{Stahl1996} measure a maximum edge velocity
$v_{\rm edge}\approx 2500$~km~s$^{-1}$.
The net result is that the confinement parameter for \tori\ is in
the approximate range $5 \le \eta_{\star} \le 20$.
The low terminal wind speed, $v_{\infty}\approx 510$~km~s$^{-1}$,
measured by \citet{Prinja1990} is lower than the escape velocity of \tori,
$v_{\rm esc} \approx 1000$~km~s$^{-1}$.
Our $\eta_{\star}\approx7.5$ model produces a $1470$~km~s$^{-1}$ wind above the
magnetic pole.

Starting from an intial time $t=0$ when the dipole magnetic field is 
suddenly introduced into a smooth, steady, spherically symmetric, CAK 
wind, our MHD simulation follows the evolution through a total time interval,
$t = 500$~ks, that is many characteristic wind flow times, 
$R_{\star}/v_{\infty} \approx 5$~ks.
Figure~\ref{f5} is a snapshot of the spatial distribution of log emission
measure per unit volume (upper panel) and log temperature (lower panel)
at a time, $t=375$~ks, when material is falling back onto the photosphere along
magnetic field lines (solid lines).

After introduction of the field, the wind stretches open the field lines
in the outer region. In the inner reqion, the wind is
channeled toward the magnetic equator by the closed field lines.
Within these closed loops, the flow from opposite hemisphere collides to 
make strong, X-ray emitting shocks. At early times, the shocks are nearly
symmetric about the magnetic equatorial plane, like those predicted in the
semi-analytic, fixed-field models of \citet{BM1997b}.
2D movies\footnote{http://www.bartol.udel.edu/t1oc} of \tori\ 
show the evolution of density, temperature, and wind speed from the
initially dipolar configuration.

Fig. \ref{f5} shows, however, that the self-consistent, dynamical 
structure is actually much more complex.
This is because once shocked material cools, its support 
against gravity by the magnetic tension 
along the convex field lines is inherently unstable, leading 
to a complex pattern of fall back along the loop lines down to the star.
As this infalling material collides with the outflowing wind driven 
from the base of a given loop line, it forms a spatially complex, time-variable 
structure, thereby modifying the location and strength of the X-ray emitting 
shocks. When averaged over time, however, the X-ray emission appears to be 
symmetric and smooth.

It should be emphasized here that the implied time variability of 
shock structure and associated X-ray  emission in this simulation is made 
somewhat artificial by the assumed 2-D, axisymmetric nature of the model.
From a more realistic 3-D model (even without rotation), 
we expect the complex infall structure would generally 
become incoherent between individual loop lines at different azimuths 
(i.e. at different magnetic longitudes). 
As such, the azimuthally averaged, X-ray emission in a
realistic 3-D model is likely to be quite similar to the time-averaged 
value in the present model.
A reliable determination of the expected residual level of observed X-ray 
variability in such wind confinement models must await future 3-D simulations 
that account for the key physical processes setting the lateral coherence scale 
between loops at different azimuth.
But based on the fine scale of magnetic structure in resolved systems
like the solar corona, it seems likely that this lateral coherence scale 
may be quite small, implying then only a low level of residual 
variability in spatially integrated X-ray spectra.

Overall, we thus see that introduction of the initial dipole magnetic field
results in a number of transient discontinuities that quickly die away.
The wind in the polar region stretches the magnetic field into a nearly radial
configuration and remains quasi-steady and smooth for the rest of the simultion.
However, near the equatorial plane where the longitudinal component of the
magnetic field, $B_{\theta}$, is large, the wind structure is quite complex
and variable.  
In the closed field region within an Alfv\'en radius, $R\approx 2R_{\star}$,
the field is nearly perpendicular to the magnetic equatorial plane and strong
enough to contain the wind. The material from higher latitudes is thus
magnetically channeled towards the equator and can reach velocities
$v_\theta\approx 1000$ km~s$^{-1}$. 

Such high speeds lead to strong equatorial shocks, with post-shock gas
temperatures of tens of millions of degree within an extended cooling layer. 
The resulting cool, compressed material is then too dense to be 
supported by radiative driving, and so falls back onto
the surface in a complex ``snake''-like pattern (see Fig.~\ref{f5}).
The equatorial material further away from the 
surface gains the radial component of momentum from the channeled 
wind, eventually stretching the field lines radially.
Eventually, the fields reconnect, allowing this material to break out. 
This dynamical situation provides two natural mechanisms for emptying the
magnetosphere: gravitational infall onto the stellar surface and break-out
at large radii.

A key result here is that the bulk of the shock-heated plasma in the
magnetosphere is moving slowly, 
implying very little broadening in the resulting X-ray emission lines.
Most of the X-rays are produced in a relatively small region
around $R\approx2R_\star$ where the density of hot gas is high.
Although the infalling and outflowing components are moving faster,
much of the infalling material is too cool to emit X-rays, while
the outflowing plasma's density is too low to produce very much
X-ray emission.

To quantify these temperature and velocity characteristics, 
we post-processed the results of these
MHD simulations to produce an emission measure per unit volume
at each 2D grid point. Then, accounting for viewing geometry and occultation
by the stellar photosphere, emission line profiles and broad-band X-ray spectra
were generated for a number of viewing angles.
In this procedure we account for the effects of geometry and occultation
on line shapes and fluxes, but we neglect attenuation by
the stellar wind and the dense equatorial outflows.
While there is empirical evidence of some excess absorption in the
magnetic equatorial plane (see \S5.3),
the effect is much smaller than would be expected
from a dense equatorial cooling disk, as in the \citet{BM1997b} model.
Our MHD simulations indicate that an X-ray absorbing cooling disk never
has a chance to form because of infall onto the photosphere, and 
outflow along the magnetic equatorial plane.

Figure~\ref{f6} plots the volume emission-measure distribution per
$\log T=0.1$ bin for the snapshot in Fig.~\ref{f5}.
The MHD simulation is used to compute the emission-measure per unit volume
per logarithmic temperature bin, then integrated over 3D space
assuming azimuthal symmetry about the magnetic dipole axis.

The resulting synthesized X-ray
emission lines are quite narrow, $\xi \leq 250$~km~s$^{-1}$, 
with symmetric profiles that are only
slightly blueshifted, $-100 \leq v_{\rm r} \leq 0$~km~s$^{-1}$.
They are thus quite distinct from profiles expected from a wind outflow
\citep{KramerRSI2003}. 
The regions of highest emission measure are located
close to the star, $R\approx2R_\star$;
the total volume emission measure
(visible at rotational phase 0.0) exceeds $10^{56}$~cm$^{-3}$;
the eclipse fraction is 20--30\%;
and the temperature distribution peaks at 20--30~MK, depending on the
simulation snapshot. 
The resulting narrow line profiles and hard
X-ray spectra are in marked contrast with observed X-ray properties
of supergiant O stars like
\zpup\ \citep{Kahn2001,OC2001,Ignace2001,IG2002,Kramer2003}.
But they match quite well the \chandra\ spectra of
\tori\ reported in this paper and of a handful of young OB stars
like $\tau$~Scorpii \citep{Cohen2003} and the other components of $\theta$~Ori
\citep{Schulz2003}.

\section{X-ray Spectral Diagnostics}

In the context of the picture of \tori\ that we have painted in the
last two sections --- that of a young hot star with a strong wind and a tilted
dipole field that controls a substantial amount of circumstellar
matter --- we will now discuss the X-ray spectral diagnostics that can
further elucidate the physical conditions of the circumstellar matter
and constrain the physics of the magnetically channeled wind.  The
reduction and analysis of the four \chandra\ grating observations have
been described in \S2, and here we will discuss the diagnostics
associated with (1) the line widths and centroids, which contain
information about plasma dynamics, (2) the line ratios in the
helium-like ions, which are sensitive to the distance of the X-ray
emitting plasma from the stellar photosphere, and (3) the 
the abundance, emission-measure, temperature, and radial-velocity
variations with viewing angle, which provide additional
information about the plasma's geometry.

\subsection{Emission Line Widths and Centroids}

The Doppler broadening of X-ray emission lines provides direct
information about the dynamics of the hot plasma that is ubiquitous on
early-type stars. For these stars, with fast, dense radiation-driven
winds, X-ray line widths and profiles have been used to test the
general idea that X-rays arise in shock-heated portions of the outflowing
stellar winds, as opposed to magnetically confined coronal loops,
as is the case in cool stars.  Single O supergiants thus far observed with
\chandra\ and \xmm\ generally show X-ray emission lines that are broadened
to an extent consistent with the known wind velocities
\citep{WC2001,Kahn2001,Cassinelli2001}, namely $v_{\rm HWHM} \approx
1000$ km s$^{-1}$, which is roughly half of \vinf, as would be
expected for an accelerating wind. Very early B stars, though they
also have fast stellar winds, show much less Doppler broadening
\citep{Cohen2003}.  Some very late O stars, such as $\delta$ Ori
\citep{Miller2002}, and unusual Be stars such as $\gamma$ Cas
\citep{Smith2004}, show an intermediate amount of X-ray line
broadening. It should be noted that although short-lived bulk plasma
flows are seen during individual flare events on the Sun, cool stars do not
show highly Doppler broadened X-ray emission lines in \chandra\ or \xmm\
grating spectra. It should also be noted that
with high enough signal-to-noise, \chandra\ HETGS data can be used
to measure Doppler broadening of less than 200 km s$^{-1}$
\citep{Cohen2003}.

In addition to a simple analysis of X-ray emission line widths,
analysis of the actual profile shapes of Doppler-broadened lines can
provide information about the dynamics and spatial distribution of the
X-ray emitting plasma and the nature of the continuum absorption in
the cold wind component
\citep{MacFarlane1991,Ignace2001,OC2001,Cohen2002,Feldmeier2003,KramerRSI2003}.
These ideas have been used to fit wind-shock X-ray models to the \chandra\
HETGS spectrum of the O4 supergiant \zpup\ \citep{Kramer2003}.  This
work, and qualitative analyses of observations of other O stars,
indicates that wind attenuation of X-rays is significantly less
important than had been assumed in O stars, but that generally, the
wind-shock framework for understanding X-ray emission from early-type
stars is sound, at least for O-type supergiants.

For \tori, where the magnetic field plays an important role in
controlling the wind dynamics, the specific spherically symmetric wind
models of X-ray emission line profiles mentioned above may not be
applicable.  However, the same principles of inverting the observed
Doppler line profiles to infer the wind dynamics and effects of
cold-wind absorption and stellar occultation are, in principle,
applicable to \tori\ as well, if an appropriate quantitative model can
be found.

In their work on the magnetic field geometry of \tori,
\citet{Donati2002} calculated several quantitative line profiles based
on the (rigid-field) MCWS model and various assumptions about the
spatial extent of the magnetosphere, the wind mass-loss rate, and the
optical thickness of the ``cooling disc'' that is thought to form in
the magnetic equator. 
\citet{Donati2002}  examined the Babel \& Montmerle model
for five sets of stellar/disk parameters. Their Model 3
produces relatively narrow, symmetric X-ray lines (as observed),
and requires very little absorption in the cooling disk.
The location of the X-ray emitter, its emission measure, the X-ray
line profiles, and the time variability can be correctly predicted
by accounting for the interaction between the wind and the magnetic field.

Line width and profile determinations for \tori\ have
already been made based on the two GTO \chandra\ grating spectra.
Unfortunately, two divergent results are in the literature based on
the same data.  \citet{Schulz2000} claimed that the emission lines in
the HETGS data from the OBSID 3, 31 October 1999, observation were
broad (mean FWHM velocity of 771 km s$^{-1}$ with some lines having
FWHMs as large as 2000 km s$^{-1}$).  In a second paper, based on both
OBSID 3 and 4 data sets, \citet{Schulz2003} claimed that all the
lines, with the exception of a few of the longest-wavelength lines,
were unresolved in the HETGS spectra. This second paper mentions a
software error and also the improper accounting of line blends as
the cause for the discrepancy.  However, as we show below, the results
of modest, but non-zero, line broadening reported in the first paper
\citep{Schulz2000} are actually more accurate.

\citet{Schulz2003} detected 81 lines in the combined HEG/MEG spectra 
\tori\ from OBSID 3 and 4. Although 82 lines are listed here in Table~2
of this paper, 35 lines have no flux errors, signifying that they
were blended with other brighter lines. Hence, \citet{Schulz2003} detect
more lines, but at lower S/N. Generally, \citet{Schulz2003} find comparable
line fluxes to those in Table~2, because OBSID 3 and 4 were obtained
at intermediate phases and the combined HEG and MEG spectra presented
here are the sum of spectra obtained at low, intermediate and high phases.

With the detailed analysis we carried out and described in \S2, we
find finite, but not large, line widths, as indicated in Table~2.
The lines are clearly resolved, as we show in Figures~\ref{f7} and \ref{f8},
in which two strong lines, one in the HEG and one in the MEG,
are compared to intrinsically narrow models, Doppler-broadened 
models, and also to data from a late-type star with narrow emission lines.

We took a significant amount of care in deriving the line widths,
fitting all lines in a given complex and carefully accounting for
continuum emission.  We fit each of the four observations separately, 
but generally find consistent widths from observation to observation, 
as we discuss below. In Figure~\ref{f9}, the Doppler width $v_{\rm D}$
of each line (open circles) is plotted versus the peak $\log T$ of that line.
The rms turbulent velocities $\xi$ are plotted as filled diamonds.
The Doppler widths are slightly higher than 
the rms velocities because the Doppler width includes the effects
of thermal broadening.  The velocities do not appear to depend on temperature,
however there are two anomalous line profiles.  
The \ion{O}{8} Lyman-$\alpha$ at 18.97~\AA\ ($\xi\approx 850$~km~s$^{-1}$)
and \ion{Fe}{17} at 15.01 \AA\ ($\xi\approx 700$~km~s$^{-1}$)
are substantially broader than any of the other lines.  Excluding 
these two lines, we find a mean $v_{\rm D} = 351 \pm 72$ km s$^{-1}$ and 
a mean $\xi=345\pm88$~km~s$^{-1}$. Because \ion{O}{8} and \ion{Fe}{17}
tend to form at lower temperatures ($T \approx 5\times10^6$)
the two broader lines could represent cooler plasma
formed by the standard wind instability process. Approximately 80\% of
the total emission measure is in the hotter 30~MK plasma. 
For lines formed in this hotter plasma, the Doppler broadening is substantially
lower than the wind velocity. The modeled line widths from the MHD
simulations, however, are quantitatively consistent with the results derived
from the data.

The line widths we derive from the data are relatively, but
not totally, narrow, and they are also approximately constant as a function of
rotational phase and magnetic field viewing angle.  We show
these results in Figure~\ref{f10}.  This is somewhat
surprising in the context of the MCWS model as elucidated by
\citet{BM1997b} and \citet{Donati2002}, where the pole-on viewing
angle is expected to show more Doppler broadening, as the shock-heated
plasma should be moving toward the magnetic equatorial plane along the
observer's line of sight. However, as we showed in \S4, the full
numerical simulations predict non-thermal line widths of 100--250~km~s$^{-1}$.
In Fig.~\ref{f10} we show the Doppler velocity from a simulation when
the magnetosphere was stable (cf. Fig.~\ref{f5}) as gray filled diamonds.

Finally, the dynamics of the shock-heated plasma in the MCWS model
might be expected to lead to observed line centroid shifts with phase,
or viewing angle.  This would be caused by occultation of the far side of
the magnetosphere by the star or by absorption in the
cooling disk (see \citet{Donati2002}).  

As predicted by the MHD simulations, the data show line centroids very close
to zero velocity at all viewing angles. Note, though,
the global fits discussed in \S5.3 show evidence
for a small viewing-angle dependence.

\subsection{Helium-like Forbidden-to-Intercombination Line Ratios}

The ratio of forbidden-to-intercombination ($f/i$) line strength in
helium-like ions can be used to measure the electron density of the X-ray
emitting plasma and/or its distance from a source of FUV radiation,
like the photosphere of a hot star \citep{Blumenthal1972}. Figure~\ref{f11}
is an energy-level diagram for \ion{S}{15} similar to the one for
\ion{O}{7} first published by \citet{GJ1969}. It shows the energy levels (eV)
and transition wavelengths (\AA) from APED of the
resonance ($^1{\rm P}_1 \rightarrow ~^1{\rm S}_0$),
intercombination ($^3{\rm P}_{2,1} \rightarrow ~^1{\rm S}_0$), and
forbidden ($^3{\rm S}_1 \rightarrow ~^1{\rm S}_0$) lines. Aside from the
usual ground-state collisional excitations (solid lines) and radiative decays
(dashed lines), the metastable $^3{\rm S_1}$ state can be depopulated by
collisional and/or photoexcitation to the $^3{\rm P}$ states.
In late-type stars, the $f/i$ ratios of low-Z ions like \ion{N}{6} and
\ion{O}{7} are sensitive to changes in electron density near or above
certain critical densities (typically in the $10^{12}$~cm$^{-3}$ regime).
Higher-Z ions have higher critical electron densities usually not
seen in normal (non-degenerate) stars.
\citet{Kahn2001} have used the $f/i$ ratios of \ion{N}{6}, \ion{O}{7},
and \ion{Ne}{9} and the radial dependence of the incident UV flux to determine
that most of the emergent X-rays from the O4 supergiant $\zeta$~Pup
were formed in the far wind, consistent with radiately driven wind shocks.

In our observations
of \tori, the He-like lines of \ion{Fe}{25}, \ion{Ca}{19},
\ion{Ar}{17}, \ion{S}{15}, \ion{Si}{13}, and \ion{Mg}{11}, were 
detected at all phases. However, accurate $f/i$ ratios cannot be 
determined for all ions.  The resonance and intercombination lines of 
\ion{Fe}{25} at 1.850 \AA\ and \ion{Ca}{19} at 3.177 \AA\ are blended 
in the HEG, and weak or non-existent in the MEG.  Similarly, \ion{Ne}{9} 
at 13.447 \AA\ and \ion{O}{7} at 21.602 \AA\ are weak or non-existent 
in the HEG.

Derived forbidden-to-intercombination ($f/i$) and
forbidden-plus-intercombination-to-resonance $(f+i)/r$ ratios are
listed in Table 4.  The \ion{Mg}{11}, \ion{Ne}{9},
and \ion{O}{7} forbidden lines are completely suppressed via FUV
photoexcitation of the $^3{\rm S}_1$ upper level,
while the forbidden lines of \ion{Si}{13}, \ion{S}{15}, and
\ion{Ar}{17} are partially suppressed. Fig.~\ref{f12} shows
the He-like line complexes of \ion{Ar}{17}, \ion{S}{15}, \ion{Si}{13},
\ion{Mg}{11} with their best fit models. The forbidden lines
of \ion{O}{7} and \ion{Ne}{9} (not shown in Fig.~\ref{f12})
are completely suppressed.

The \ion{Ar}{17} forbidden line at 3.994 \AA\ is blended with
\ion{S}{16} Lyman-$\beta$ at 3.990 and 3.992 \AA\ and \ion{S}{15} at
3.998 \AA. Based on the ISIS abundances and emission-measures
(see \S 5.3 below),
APED predicts that only 26\% of the HEG line flux comes from the
\ion{Ar}{17} $f$ line. In the above case, the $f/i$ ratio is highly
uncertain and strongly model dependent, making the use of this line
ratio problematic. The \ion{Si}{13} $f$ line at 6.740 \AA\ is
blended with the \ion{Mg}{12} Ly$\gamma$ doublet at 6.738 \AA.
Similarly, we estimate that the \ion{Mg}{12}
doublet's flux is approximately $11.8\times10^{-6}$~ph~cm$^{-2}$~s$^{-1}$,
approximately 38\% of the flux at 6.740 \AA. Assigning the remaining flux
to the \ion{Si}{13} $f$ line results in $f/i$ = 0.98, where we have
assigned a large 30\% uncertainty in Table~4 because of the blending problem.

This leaves us with $f/i$ measurements for \ion{Mg}{11}, \ion{S}{15},
as well as \ion{Si}{13}. We take the Mg forbidden line to be
undetected (the fit result in Table 4 shows that it is just
barely detected at the $1\sigma$ level).  The sulfur $f/i$ value is
quite well constrained, though, and lies somewhere between the two
limits (generally referred to as ``high density'' and ``low density,''
reflecting the traditional use of this ratio as a density diagnostic
in purely collisional plasmas).

Of course, the photospheric emergent fluxes must be known at the relevant UV
wavelengths for the $f/i$ ratio to determine the location of the X-ray
emitting plasma.
The \ion{Mg}{11}, \ion{Si}{13} and \ion{S}{15} photoexcitation wavelengths
are listed in Table~4.
The 1034~\AA\ flux has been measured using archival {\it Copernicus} spectra.
For the fluxes short of the Lyman limit at 912~\AA\ where we have no data and
for which non-LTE model atmospheres and synthetic spectra have
not been published, we scaled the flux from a 45000 K blackbody
(the hot model in Table~3).
We computed the $f/i$ ratios using the PrismSPECT
non-LTE excitation kinematics code \citep{MacFarlane2003}.
For each line complex, we set up a model atom with several dozen levels for
each ion using oscillator strengths and transition rates custom-computed
using Hartree-Fock and distorted wave methodologies. The
resulting values were checked against data from the literature. We include
collisional and radiative coupling among most levels in the model atom,
and specifically include photoexcitation between the $^3{\rm S_1}$
and $^3{\rm P}$ levels driven by photospheric radiation.

The results of this modeling for \ion{Mg}{11} and \ion{S}{15} are
shown in Figure~\ref{f13}, where the solid lines represent $f/i$ as a
function of electron density for a range of radii from 1--200 $R_{\star}$.
The dashed line in Fig.~14a is the \ion{Mg}{11} upper limit, indicating
an upper limit to the formation radius $R \le 1.7 R_\star$.
The dashed lines in Fig.~14b are the \ion{S}{15} $1\sigma$ upper and
lower bounds, indicating a formation radius in the range
$1.2 R_\star \le R \le 1.5 R_\star$.
The resulting bounds from \ion{Si}{13} are consistent with \ion{S}{15}.
Of course, it is a simplification to
assume that all of the plasma (even for a given element or ion stage)
is at a single radius.  The formation radii we have derived should
thus be thought of as an emission-measure weighted mean. 
In any case, it is clear
that the hot plasma in the circumstellar environment of \tori\ is
quite close to the photosphere. We note that there is no variation in
the average formation radius among the four separate phase observations. 
We note that these results are comparable to those found by \citet{Schulz2003}
based on the ealier two data sets.

\subsection{Emission-Measure and Abundance Variability}

There are several additional diagnostics that involve analysis not of
the properties of single lines, but of the X-ray spectrum as a whole.  The
global thermal model fitting with ISIS described in \S2 yields
abundance and emission-measure information, while the line flux values
taken as an ensemble can be analyzed for phase, or viewing angle, dependence.
Detailed descriptions of these analyses have already been presented \S2,
so here we simply present the results.

The elemental abundances derived from the two different fitting
methods are reported in Table 5. Most elemental abundances do not deviate
significantly from solar, although Fe is well below solar and several
intermediate atomic number elements are slightly above solar.

The HEG/MEG spectra for each observation were fit with a variable-abundance, 
2-temperature ISIS plasma model with radial velocity and turbulent
(a.k.a., rms) velocity as free parameters (See Table 6),
with a single column density parameter, $N_{\rm H}$.
We tried fitting the individual and combined spectra with 2-, 5-, and
21-temperature APEC models with solar and non-solar abundances.
No solar-abundance model can adequately fit the continuum and emission lines.
The 5- and 21-temperature component models do not provide appreciably
better fits than the 2-temperature non-solar abundance model.
Although plasma at a number of temperatures may be present, 
it is not possible to accurately fit so many free parameters with
moderate S/N HETG data.

\citet{Schulz2003} used a six-temperature, variable abundance APEC plasma
model to fit the combined HEG/MEG spectra from OBSID 3 and 4.
They find $N_{\rm H}\approx 5\times 10^{21}$~cm$^{-2}$,
sub-solar Fe, Ni, O and Ne and approximately solar abundances of
other elements, again, consistent with our results in Table 6.
Their Figure~7, however, suggests a phase dependence in the six-temperature
emission-measure distribution, which we cannot confirm. We point out,
however, that the $1\sigma$ error envelope in their emission-measure
distributions overlap significantly. The emission measures in Table~6 for
phases 0.38 and 0.84 appear to fit within both EM envelopes in Fig.~7 of 
\citet{Schulz2003}.

The visible emission measure decreases
by 36\% from X-ray maximum to minimum, suggesting a substantial fraction
of the X-ray emitting region is occulted by the star at high viewing angles.
Comparison between the observed data and
the MHD simulations show the agreement is quite good, both in terms of 
the overall volume emission measure and in its temperature distribution.

As has previously been shown \citep{Gagne1997,BM1997b,Donati2002},
the phase dependence of the X-ray spectrum provides information
about the extent and location of the X-ray emission and absorption in the wind
and disk. There are no large, systematic spectral variations at the four phases
observed with {\it Chandra}. E.g., the soft X-ray portion of the MEG is not
more absorbed than the hard X-ray portion, suggesting that the variability
is not primarily caused by absorption in the disk or by a polar wind.
Rather it appears that most of the variability is caused by occultation
of the X-ray emitter by the stellar photosphere as the magnetically confined
region rotates every 15.4 days.

That said, there does a appear to be a small increase in column density
when the disk is viewed nearly edge-on.
Figure~\ref{f14} shows the ISIS parameters,
$N_{\rm H}$ (upper curve, right axis) and radial velocity
(lower curve, left axis), as a function of viewing angle.
The excess column density, $N_{\rm H}\approx 5\times10^{20}$~cm$^{-2}$,
would arise as a result of increased outflow in the magnetic equatorial plane.
Because increased column density and increased temperature harden the
predicted X-ray spectrum in somewhat similar ways, we consider the
column density variablity as a tentative result.
We point at that X-ray absorption towards \tori\ occurs in the ISM, in
the neutral lid of the Orion Nebula, in the ionized cavity,
and in the stellar wind. The Wisconsin absorption model for
cold interstellar gas yields a single column density parameter that does
not take into account the ionization fraction, temperature and abundance
of these absorption components.

The measured radial velocities at viewing angles $4\arcdeg$ (pole-on),
$39\arcdeg$, $80\arcdeg$, and $87\arcdeg$ (edge-on) were
$-75\pm10$, $-66^{+12}_{-9}$, $19^{+19}_{-18}$, and
$93^{+16}_{-13}$~km~s$^{-1}$, respectively.
Systematic line shifts were not detected in individual line
complexes (see \S5.1) because the uncertainty in the centroid velocity of
a single line is higher than its uncertainty based on a large number of
lines. For instance, the $1\sigma$ uncertainty at each OBSID in the line
centroid of \ion{Fe}{17} $\lambda15.014$ is approximately 50~km~s$^{-1}$.

The simulations predict no significant blueshifts or redshifts at any viewing
angle. As a result, the $-75$~km~s$^{-1}$ blueshift at low viewing angle and
the $+93$~km~s$^{-1}$ redshift at high viewing angle seen in Fig.~\ref{f11},
if real, may lead to further improvements of the model.
The 2D MHD simulations show that the hot plasma
is constantly moving, sometimes falling back to the photosphere in one
hemisphere or the other, other times expanding outward in the magnetic
equatorial plane following reconnection events.
Because the four \chandra\ observations were obtained months
or years apart, the radial velocity variations may 
reflect the stochastic nature of the X-ray production mechanism
or persistent flows of shock-heated plasma.

Though small, the X-ray line shifts nicely match the \ion{C}{4},
\ion{N}{5} and H$\alpha$ emission components discussed by \citet{SF2005}.
In this picture, material falling back onto the
photosphere along closed magnetic field lines produces redshifted
emission when viewed edge-on.
The blueshifted X-ray and H$\alpha$ emission seen
at low viewing angles (pole-on, $\phi=0$) would be produced by 
(1) the same infalling plasma viewed from an orthogonal viewing angle and/or
(2) by material in the outflowing wind.
\citet{SF2005} argue that a polar outflow probably cannot produce
the observed levels of H$\alpha$ and \ion{He}{2} $\lambda4686$ emission,
thereby preferring the infall scenario.

If pole-on blueshifts in the X-ray lines are persistent, this
suggests asymmetric infall towards the north (near) magnetic pole.
Such blueshifts could be caused by an off-centered oblique dipole.
The blueshifts might also be caused by absorption of the red emission in
the far hemisphere by the cooler post-shock gas. However, this would produce
weaker X-rays at pole-on phases.
Since stronger X-ray emission is observed at pole-on phases,
asymmetric infall provides a better explanation for the blueshifted
H$\alpha$ and X-ray line emission. Repeated observations at pole-on and edge-on
phases are needed to confirm this result.

\section{Discussion}

The high-resolution X-ray spectroscopy covering the full range of
magnetospheric viewing angles adds a significant amount of new
information to our understanding of the high energy processes in the
circumstellar environment of \tori, especially when interpreted in
conjunction with data from other wavelength regions and our new MHD
numerical simulations. As the simplified MCWS modeling
\citep{BM1997b,Donati2002} indicated and our detailed MHD simulations
confirmed, the magnetically channeled wind model can produce the right
amount of X-rays and roughly the observed temperature distribution in
the X-ray emitting plasma.  The two new \chandra\ observations we
report on in this paper reinforce the global spectral information
already gleaned from the two GTO observations. In addition, the better
phase coverage now shows that the modulation of the X-rays is quite
gray, indicating that occultation by the star at phases when our view
of the system is in the magnetic equatorial plane, is the primary
cause of the overall X-ray variability.  Incidentally, the fact that
the reduction in the flux is maximal at $\phi = 0.5$ confirms the
reassessment of field geometry by \citet{Donati2002}.

The fractional decrease in the observed flux of roughly 35\%,
interpreted in the context of occultation, indicates that the X-ray
emitting plasma is quite close to the photosphere:
$1.2 R_\star \le R \le 1.4 R_\star$.
This result is completely consistent with
the observed \ion{Mg}{11}, \ion{Si}{13} and \ion{S}{15} $f/i$ values,
which together indicate $1.2 R_\star \le R \le 1.5 R_\star$,
i.e., between 0.2 and 0.5 stellar radii above the photosphere.
The MHD simulations we have performed also show the bulk of the
very hot emission measure to be, on average, within $2R_\star$.

We see evidence for slightly enhanced attenuation
at large viewing angles (nearly edge-on) as a result of wind material
channeled into the magnetic equatorial plane. We should point out
that the infall (in the closed-field region) and outflow (in the open-field
region) means that a dense, X-ray absorbing, cooling disk does not form
in the equatorial plane.

The X-ray emission lines seen in the \chandra\ spectra are well resolved,
but relatively narrow, and show small ($< 100$~km~s$^{-1}$) centroid
shifts. Although the 2D MHD simulations indicate little or no line shifts
on average, the small observed shifts (also seen in H$\alpha$) may
reflect the stochastic nature of the infall onto the star. If the
blueshifts at pole-on phases are persistent, they may indicate 
an asymmetry in the magnetic/wind geometry.

The line widths in the simulations are even narrower than those seen
in the data.  But this is, perhaps, not too surprising, as the
simulations do not include the deshadowing instability that can lead
to shock heating in the bulk wind. Some radiatively-driven X-ray production
in the cooler ($T \approx 6\times10^6$ K) wind shocks was suggested by
\citet{Schulz2003} and may explain the
larger widths of the \ion{Fe}{17} and \ion{O}{8} lines.  Furthermore,
fully three dimensional simulations may reveal
additional turbulence or other bulk motions.

\section{Conclusions}

The magnetically channeled wind shock model for magnetized hot stars
with strong line-driven winds provides excellent agreement with the
diagnostics from our phase-resolved \chandra\ spectroscopy of \tori.
The modest line widths are consistent
with the predictions of our MHD simulations. The X-ray light curve and
He-like $f/i$ values indicate that the bulk of the X-ray emitting
plasma is located at approximately $1.5 R_\star$, very close to the photosphere,
which is also consistent with the MHD simulations of the MCWS mechanism.
The simulations also correctly predict the temperature
and total luminosity of the X-ray emitting plasma.
We emphasize that the only
inputs to the MHD simulations were the magnetic field strength, 
effective temperature, radius, and mass-loss rate of \tori, all
of which are fairly well constrained by observation.

Although the original explorations of the MCWS
mechanism indicated that there might be significant viewing angle
dependencies in some of these diagnostics, the numerical simulations
by and large do not predict them, and they are generally not seen in
the data. Spatial stratification of the very hottest plasma might
explain the slight wavelength dependence of the viewing angle
variability of the X-ray flux.

The new MHD simulations we present for \tori\ are similar to the
original calculations by \citet{UO2002}, but the more accurate
treatment of the energy equation makes for some modest changes, and
allows for a direct prediction of its X-ray emission properties.
Future, fully 3D simulations might show even better agreement
with the data if different longitudinal sectors of the magnetosphere
exhibit independent dynamics. But the addition of rotation into the
dynamical equations of motion probably is not important for this star
and seems to be unnecessary for reproducing the observations.  

The red and blue wings of the \ion{C}{4} and 
\ion{N}{5} UV resonance-lines, the optical \ion{H}{1} Balmer
and \ion{He}{2} lines, and small centroid velocity shifts in the X-ray lines
may point to more complicated, episodic infall and/or mass-loss.

As was pointed out by \citet{Schulz2003}, \tori\ is one of many early-type
stars in OB associations with hard X-ray spectra.
The MCWS model may be fruitfully applied to $\tau$ Sco, and other young,
early-type stars that show relatively narrow X-ray emission lines, hard
X-ray spectra, and time variability.
  
\acknowledgements

This work was partially supported by NASA/SAO grant GO2-3024 to
West Chester University, Swarthmore College, and the University of Delaware.
DHC thanks the Howard Hughes Medical Institute for its support to
Swarthmore College. SPO and RHDT acknowledge support from NSF grant
AST-0097983 to the University of Delaware. MG thanks Myron Smith for
helping to interpret the UV and optical lines in the context of the MHD
simulations. The authors would like to thank the anonymous referee for 
suggesting many improvements to the manuscript.


\begin{figure}
\figurenum{1}
\plotone{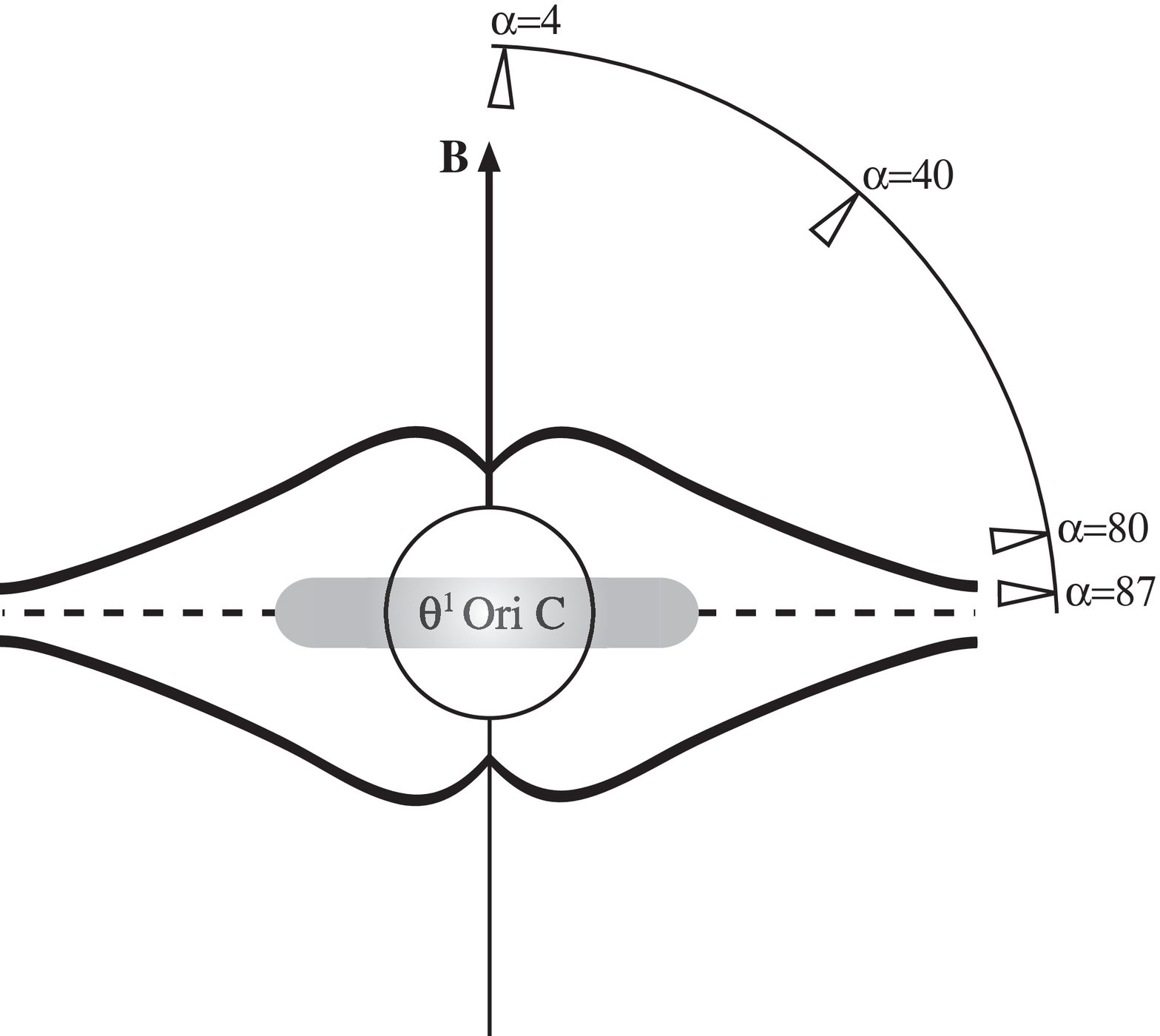}
\caption{Schematic of \tori\ and its circumstellar environment.
The curved lines represent a set of magnetic field lines stretched
out by the wind. The vector indicates the magnetic axis and the dashed line
represents the magnetic equatorial plane.
The viewing angles in degrees with respect to the magnetic axis
of the four \chandra\ observations,
$\alpha$, are indicated by the four labels, 4, 40, 80, and 87.
The arrows point to the observer.
In this figure, taken from the perspective
of the star's magnetic axis, the observer appears to move from
$\alpha\approx 3\arcdeg$ (at rotational phase 0.0) to
$\alpha\approx 87\arcdeg$ (at rotational phase 0.5),
back to $\alpha\approx 3\arcdeg$ (at rotational phase 1.0) as a result of
the magnetic obliquity, $\beta = 42\arcdeg$.
The rotation axis, which we assume is inclined by approximately $45\arcdeg$
to the observer, is not shown because it too would move.
The hard X-ray emitting region is shown schematically as a gray torus.
At high viewing angles, the entire torus is visible (X-ray maximum).
At low viewing angles, some of the torus is occulted (X-ray minimum).
\label{f1}}
\end{figure}

\begin{figure}
\figurenum{2}
\includegraphics[scale=0.85]{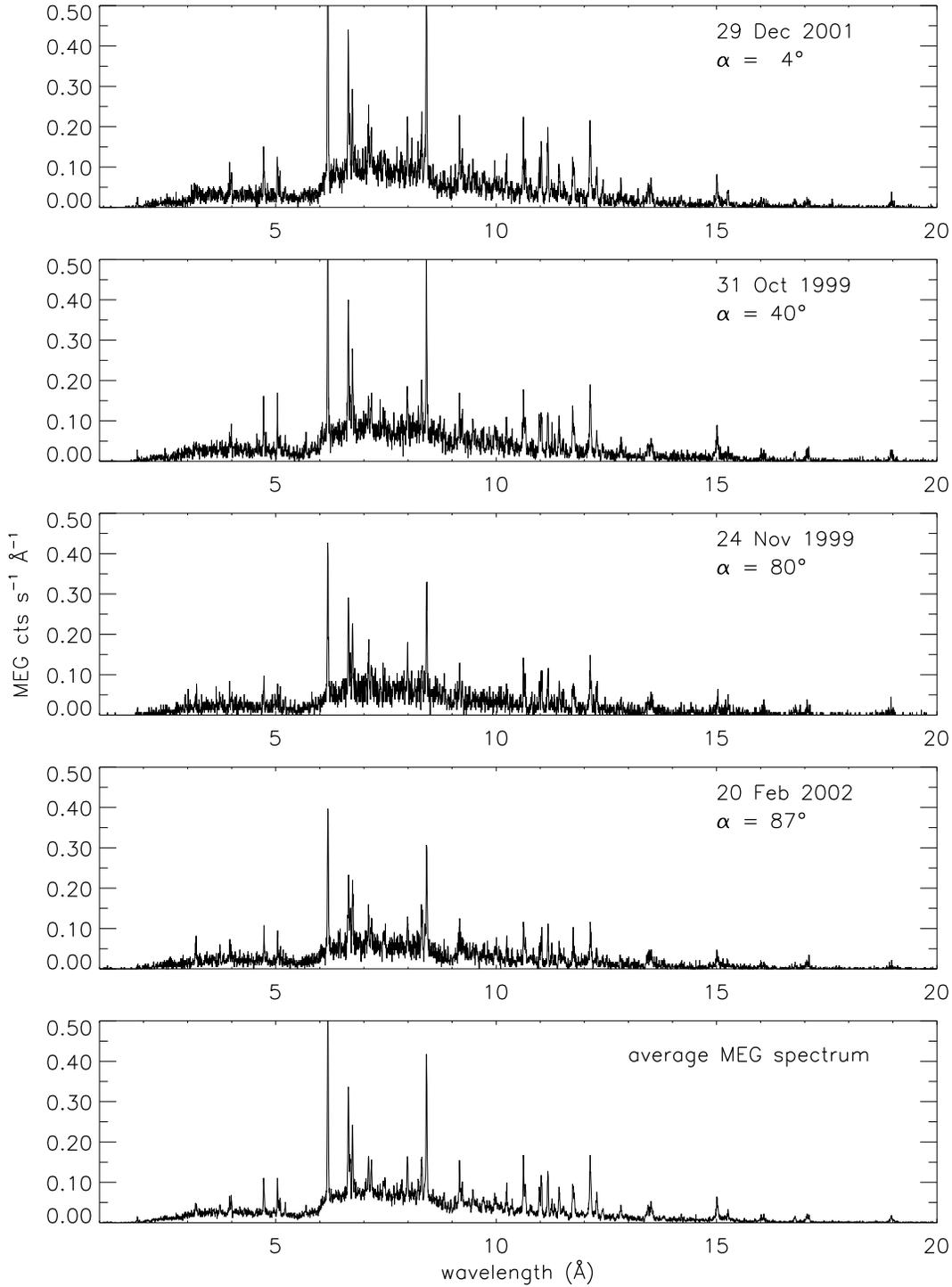}
\caption{
All four co-added, first order MEG spectra, ordered according
to viewing angle (pole-on to equator-on).  The bottom panel shows an
averaged MEG spectrum composed of the four individual spectra. The low
count rate at long wavelengths is due to ISM attenuation.
\label{f2}}
\end{figure}

\begin{figure}
\figurenum{3}
\includegraphics[scale=0.85]{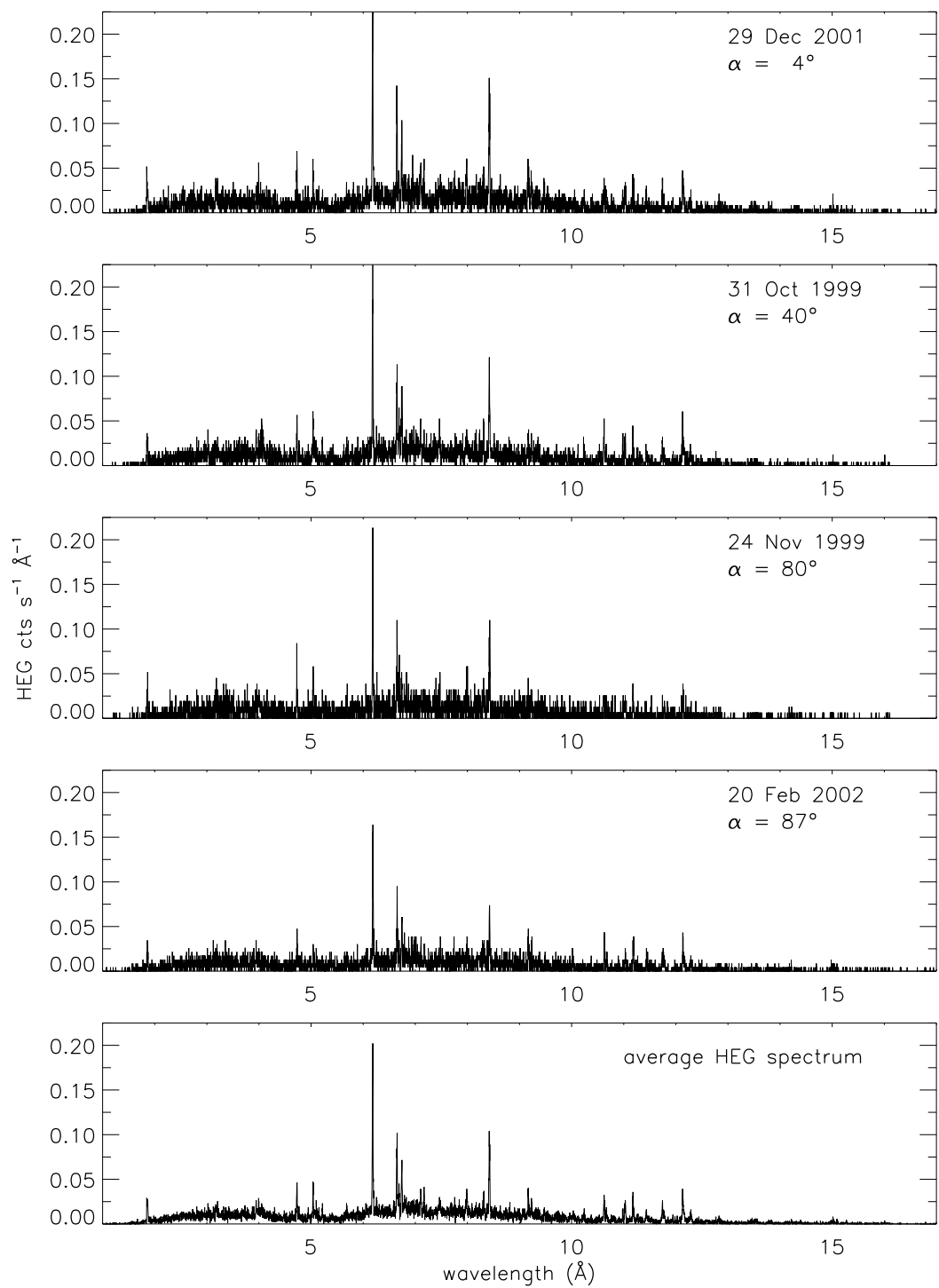}
\caption{
All four co-added, first order HEG spectra, ordered according
to viewing angle (pole-on to equator-on).  The bottom panel shows an
averaged HEG spectrum composed of the four individual spectra.
The HEG has higher spectral resolution than the MEG, but lower throughput 
longward of approximately 3.5~\AA.
\label{f3}}
\end{figure}

\begin{figure}
\figurenum{4}
\begin{center}
\includegraphics[scale=0.38]{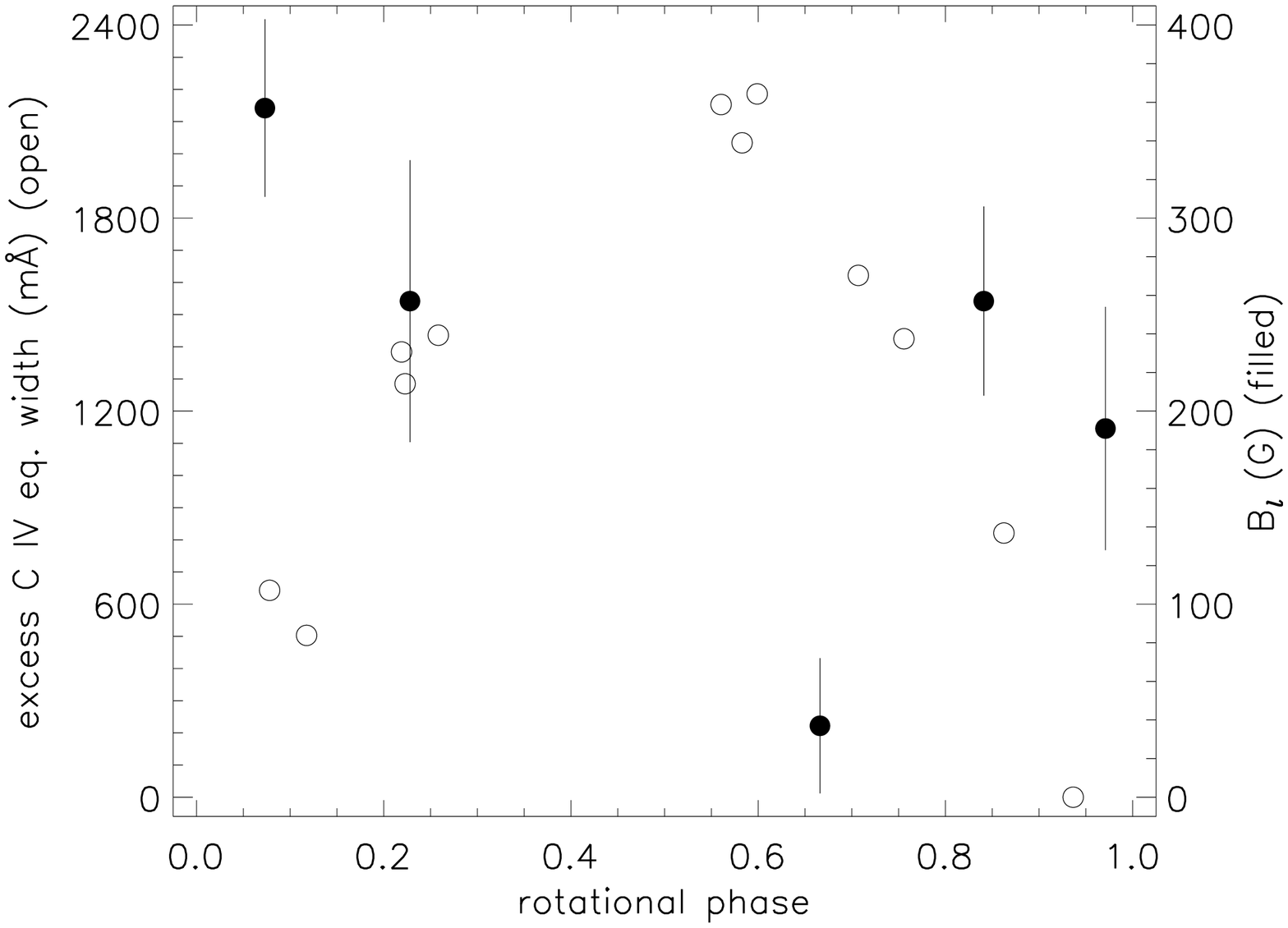}
\includegraphics[scale=0.35]{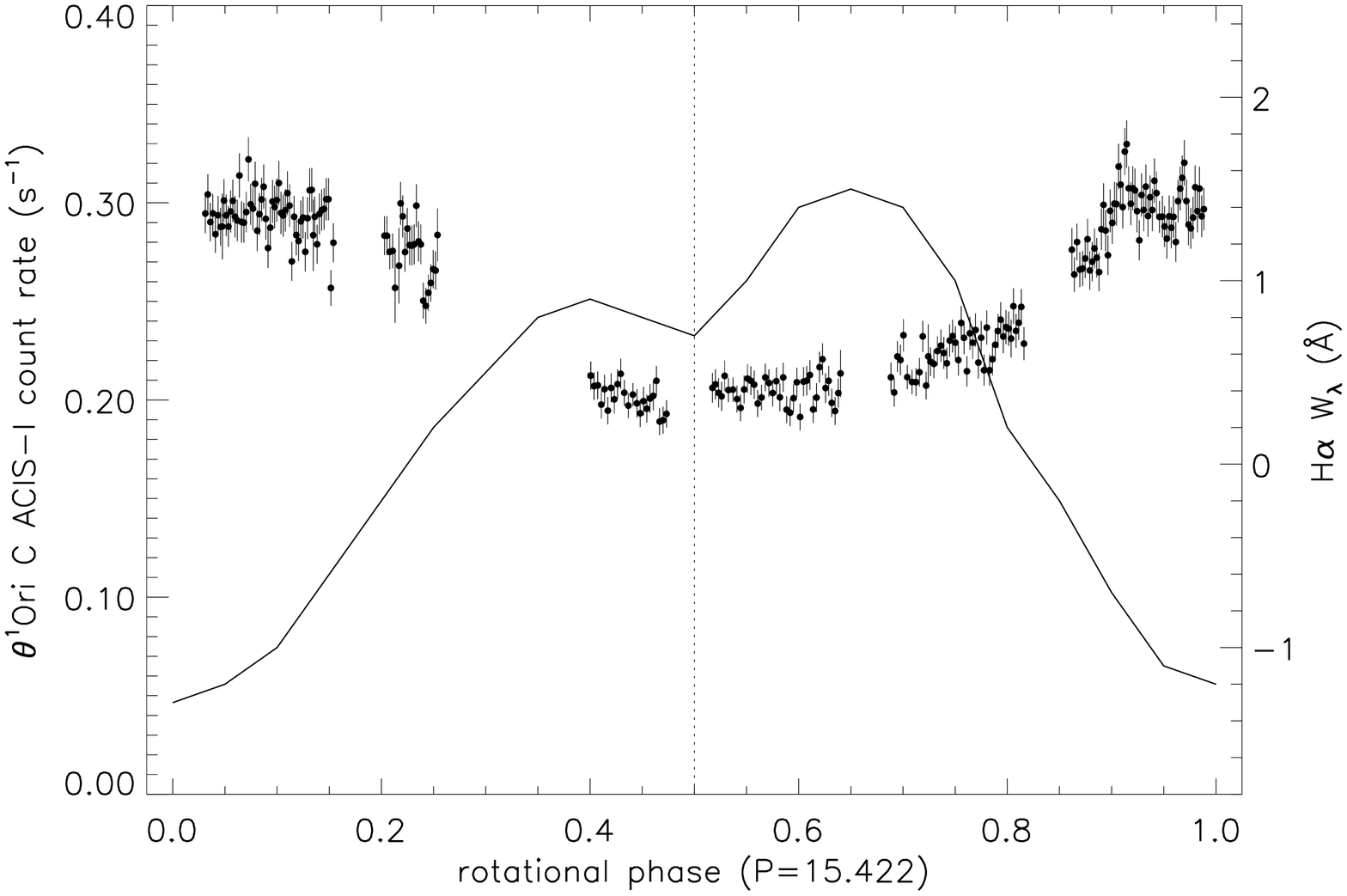}
\end{center}
\caption{
Phase-folded light curves of \tori.
Upper panel: open circles indicate the excess \ion{C}{4} equivalent width
(left axis) taken from \citet{WN1994} and phased to the ephemeris of
\citet{Stahl1996}: period $P=15.422$~days and epoch
${\rm MJD}_{0}$ = 48832.50. Maximum \ion{C}{4} absorption occurs near phase
0.5 ($\alpha=3\arcdeg$) as a result of outflowing plasma in the
magnetic equatoral plane. Note: \citet{WN1994} calculate $W_\lambda$ by 
subtracting the IUE spectrum at a given phase from the IUE spectrum
with the shallowest line profile, then calculating the equivalent width
of the line in the difference spectrum.
Filled circles show the longitudinal magnetic field strength, $B_\ell$,
(right axis) as measured by \citet{Donati2002} using the same ephemeris.
Note that \citet{WN1994} and \citet{Donati2002} used different
period estimates. $B_\ell$ is maximum near phase 0.0 when
the magnetic pole is in the line of sight.
Lower panel: H$\alpha$ equivalent width (solid curve) from \citet{Stahl1996}.
The data points with error bars represent the ACIS-I count rate
from the 850-ks {\it Chandra} Orion Ultra-Deep Project.
X-ray and H$\alpha$ maxima occur at low
viewing angles when the entire X-ray torus is visible;
the minima occur when part of the X-ray torus is occulted by the star.
\label{f4}}
\end{figure}

\begin{figure}
\figurenum{5}
\begin{center}
\includegraphics[scale=0.45]{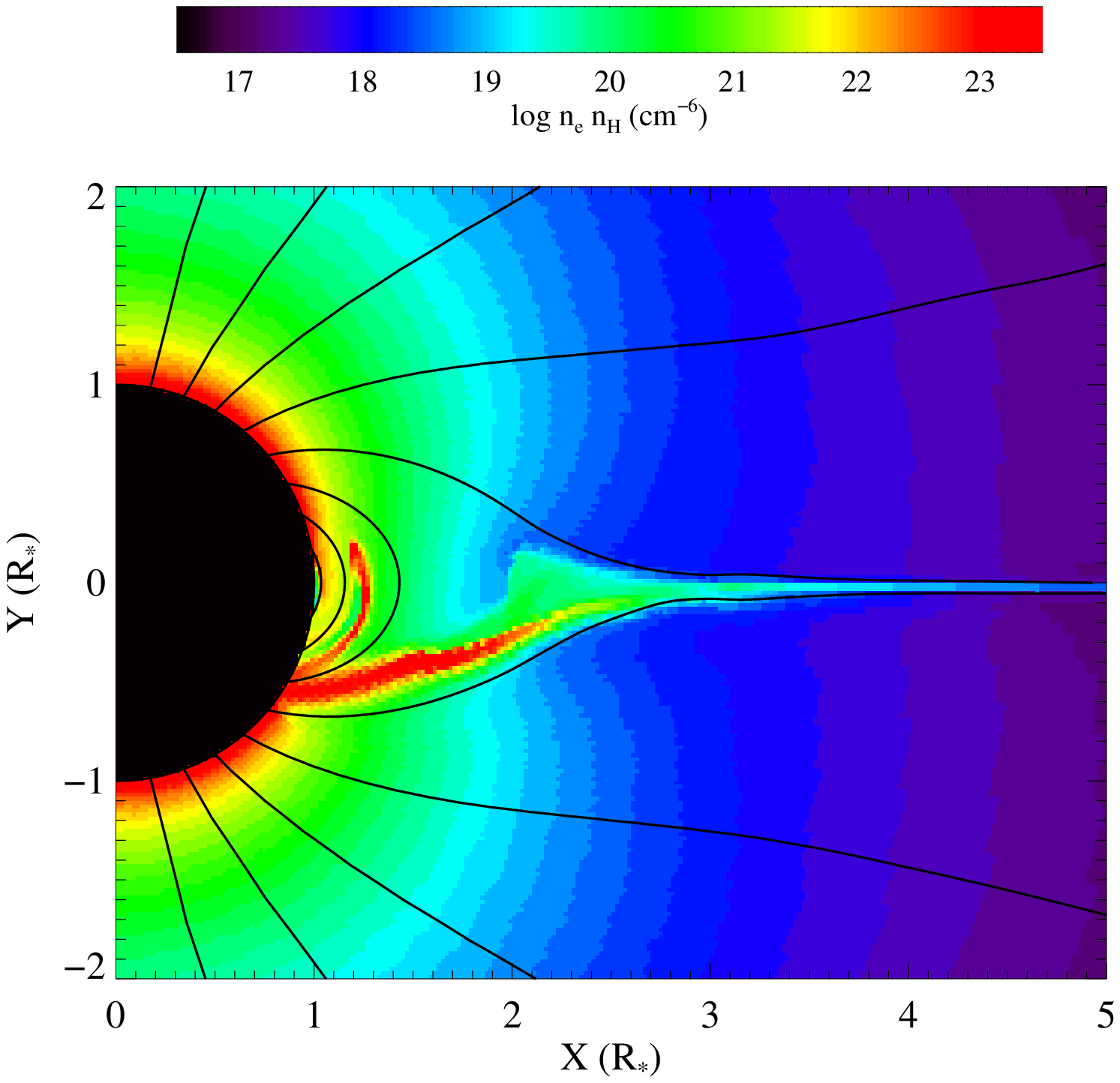}
\includegraphics[scale=0.45]{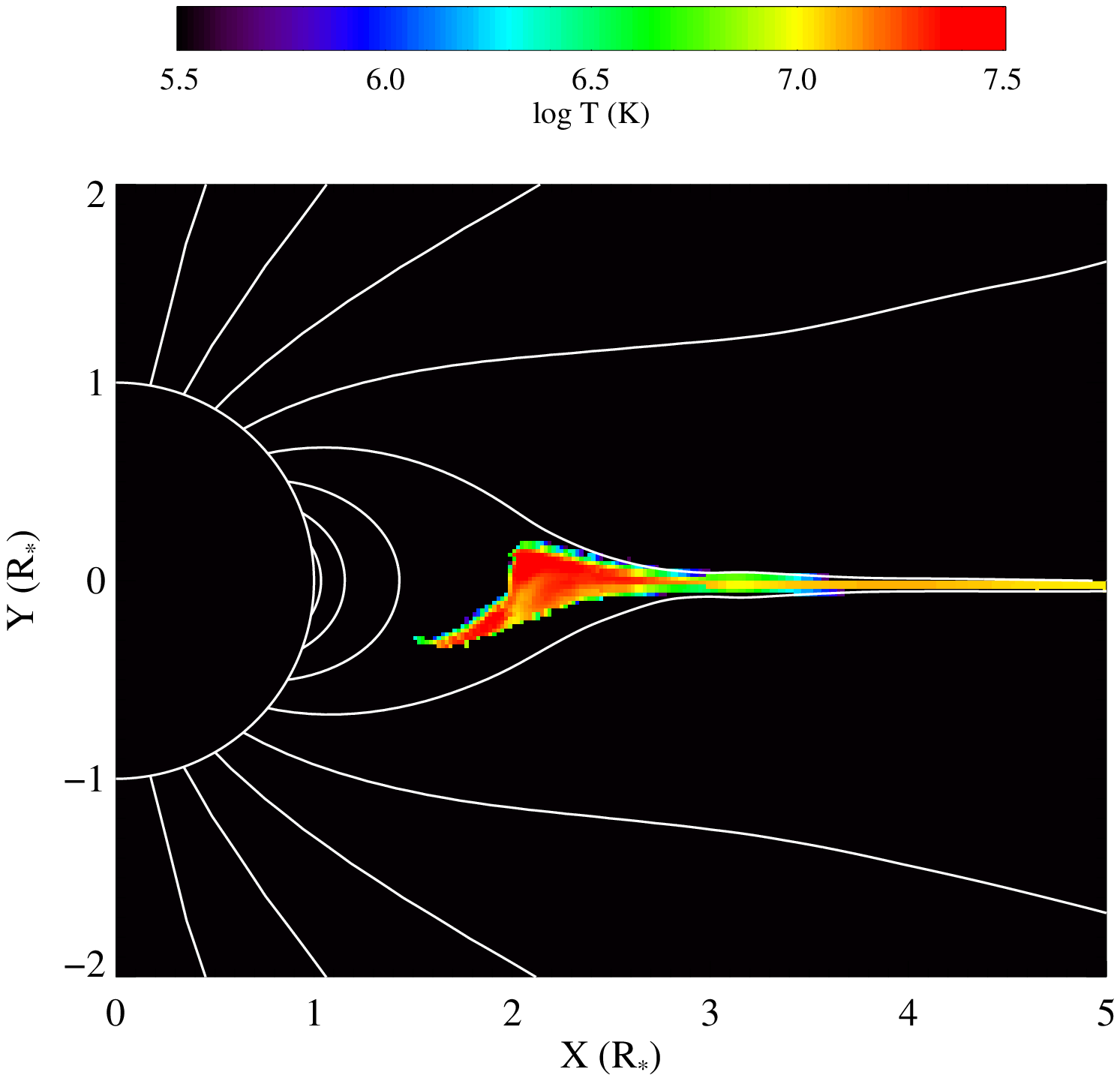}
\end{center}
\caption{
Grayscale (print edition) and color (electronic edition)
snapshots from the 500-ks, slow-wind, 2D MHD simulation of \tori, superposed
with corresponding magnetic field lines.
Upper panel: logarithm of the emission measure per unit volume,
$\log n_{\rm e} n_{\rm H}$.
Lower panel: logarithm of temperature,  $\log T$.
In this snapshot, obtained at simulation time $t=375$~ks,
material trapped in closed loops is falling back toward the stellar surface
along field lines, forming a complex ``snake-like'' pattern.
The regions of highest emission measure in the 1--100~MK temperature range
occur above and below the magnetic equator, close to $R\approx 2R_{\star}$.
\label{f5}}
\end{figure}

\begin{figure}
\figurenum{6}
\plotone{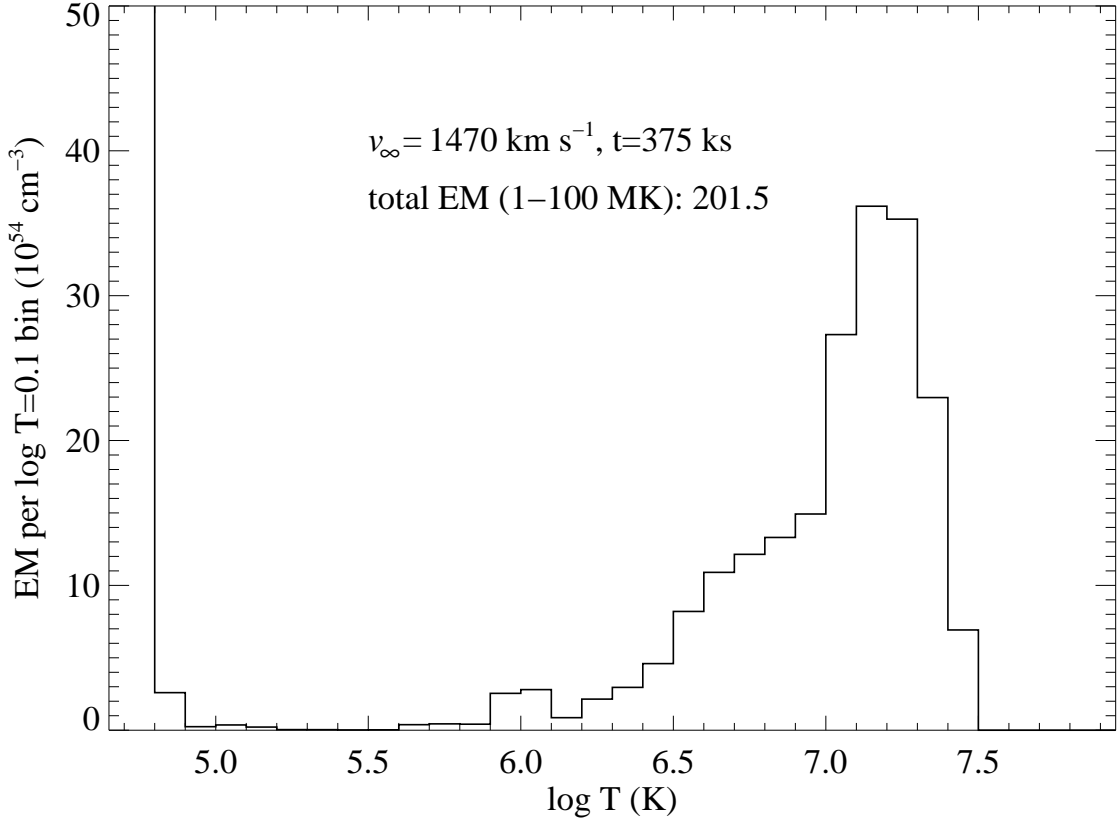}
\caption{
Volume emission-measure distribution per $\log T=0.1$ bin,
from the snapshot in Fig.~5.
The MHD simulation is used to compute the emission-measure per unit volume 
per logarithmic temperature bin, then integrated over 3D space
assuming azimuthal symmetry about the magnetic dipole axis.
The total simulated volume emission measure in the interval
$6.0 \leq \log T < 8.0$ exceeds the total emission measure seen with the HETG
by a factor of two.
The low-velocity wind ($v_\infty \approx 1470$~km~s$^{-1}$) simulation
in Fig.~5 also produces a slightly cooler distribution of shocks
than observed. Full 3D MHD simulations may be needed to resolve these
discrepancies.
\label{f6}}
\end{figure}

\begin{figure}
\figurenum{7}
\begin{center}
\includegraphics[scale=0.32]{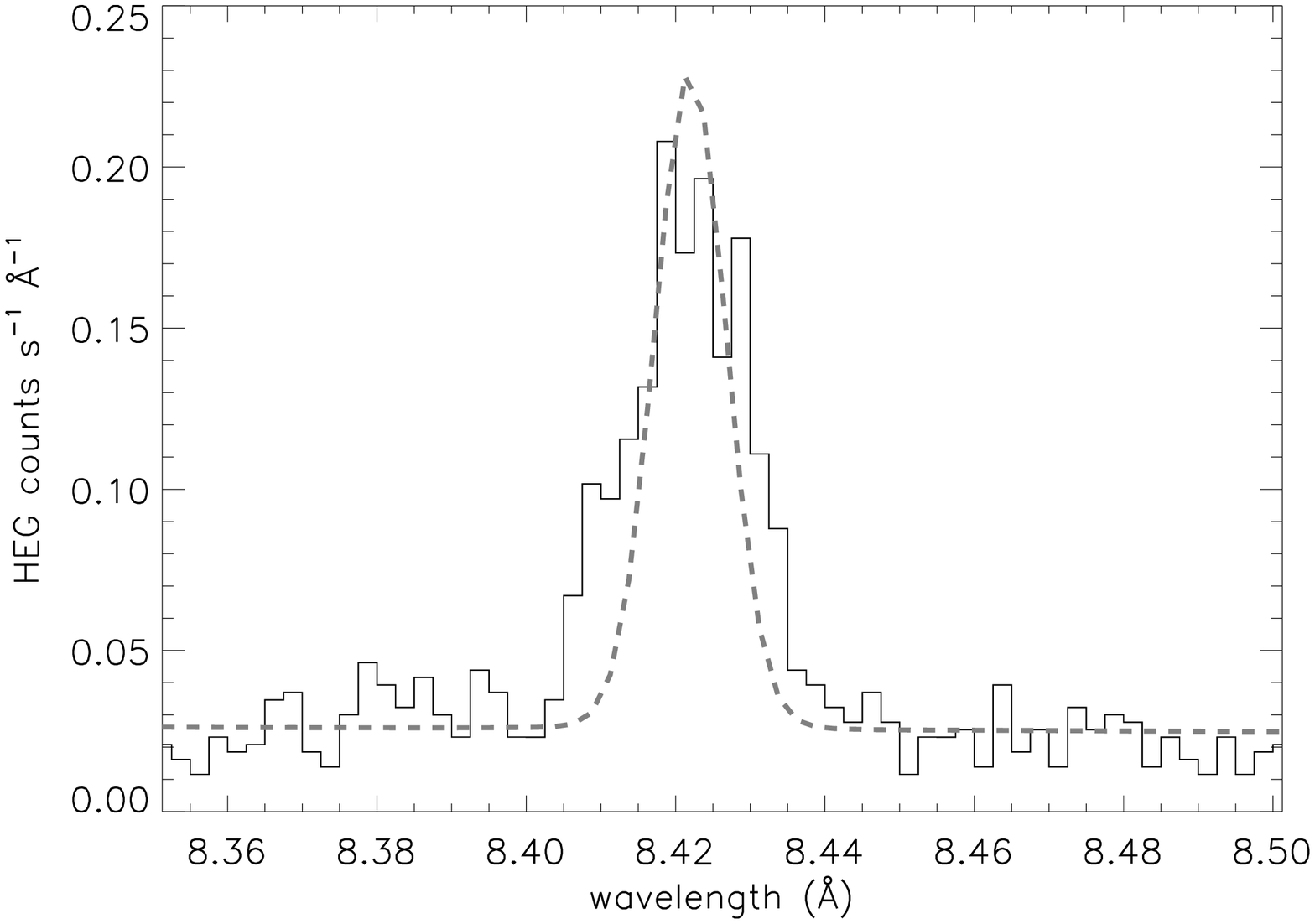}
\includegraphics[scale=0.32]{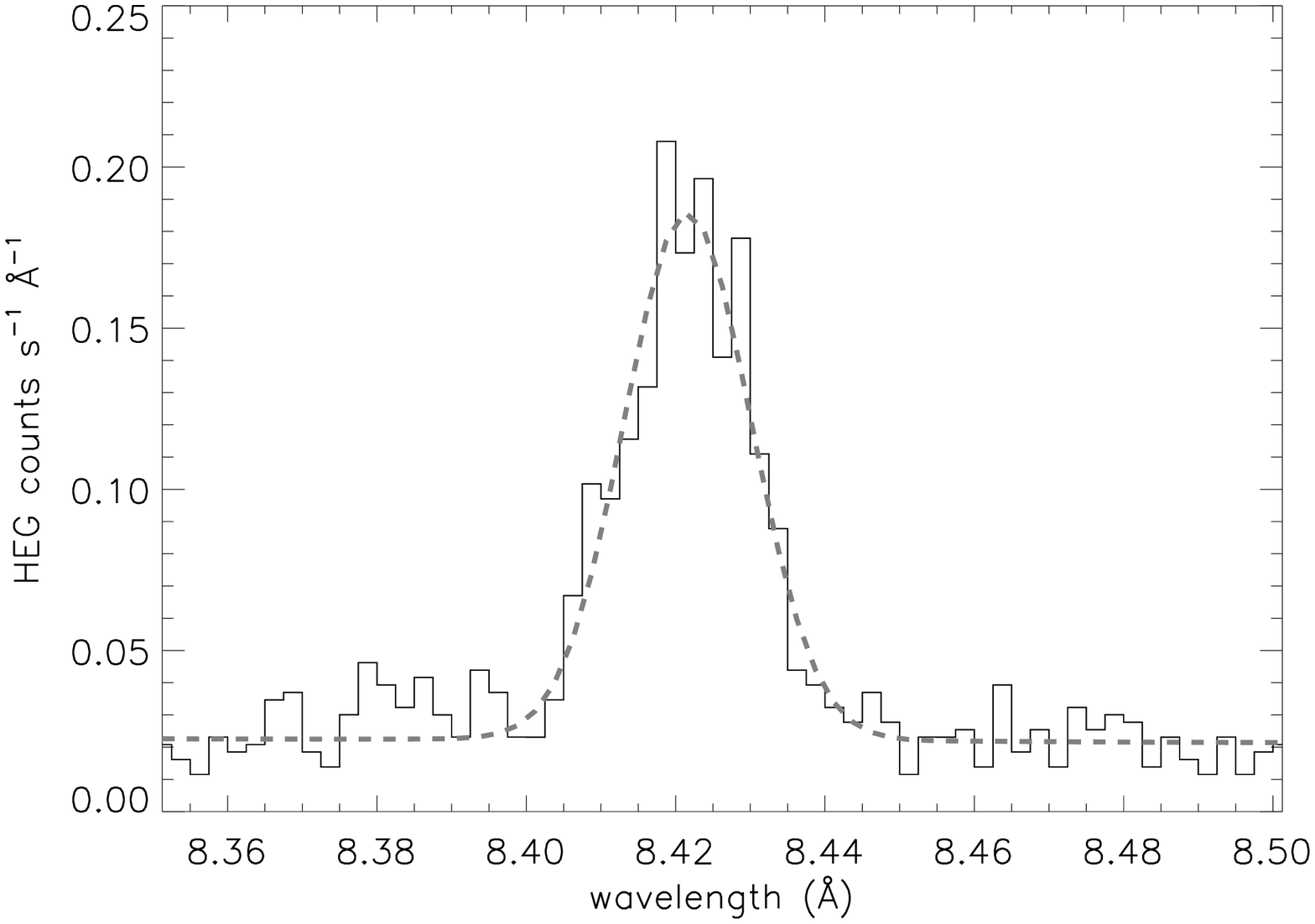}
\end{center}
\caption{
The Mg Lyman-$\alpha$ line in the combined \chandra\ HEG
spectrum from all four observations (solid histogram, both panels).
A model with only thermal broadening
($v_{\rm th} = 104$~km~s$^{-1}$; dashed line, upper panel)
is much narrower than the observed profile.
A Gaussian line model wth thermal and turbulent broadening due to bulk motion
($\xi = 328$ km s$^{-1}$; dashed line, lower panel)
provides a better fit to the data.
\label{f7}}
\end{figure}

\begin{figure}
\figurenum{8}
\plotone{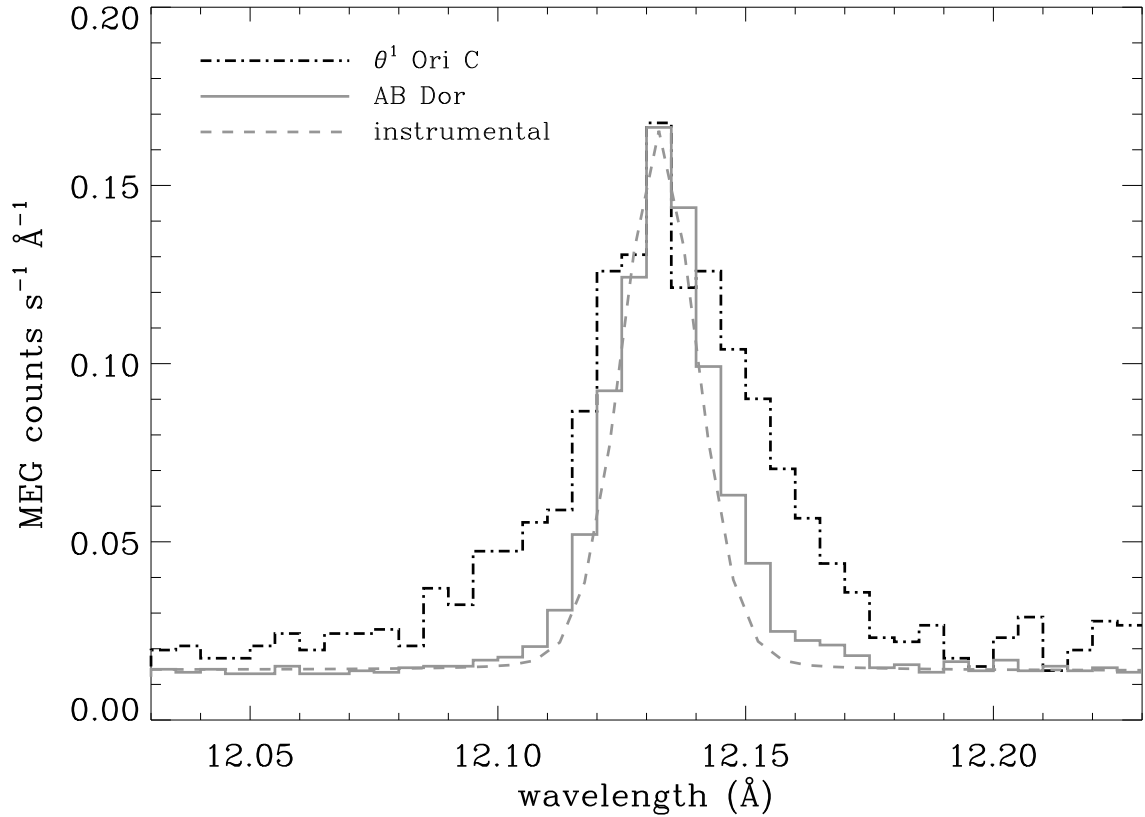}
\caption{
The Ne Lyman-$\alpha$ line in the combined \chandra\ MEG spectrum
from all four observations of \tori\ (solid histogram) compared to the same
line seen in the MEG spectrum of the active young K-type dwarf, AB Doradus
(dash-dot histogram). An delta function convolved with the MEG
instrumental response (dashed line) is also shown for comparison.  
The \tori\ line is clearly broader than both the narrow line or the AB Dor line.
\label{f8}}
\end{figure}

\begin{figure}
\figurenum{9}
\plotone{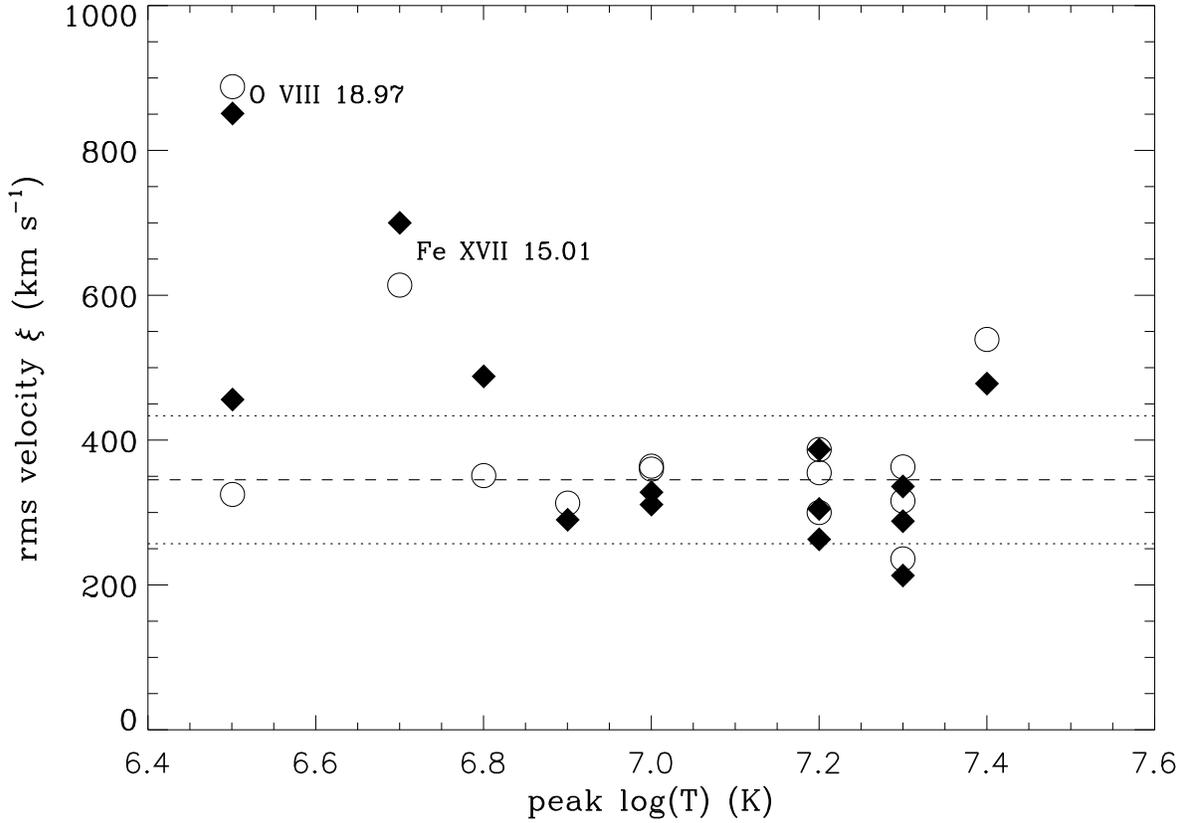}
\caption{
Line widths for the strongest lines in the \chandra\ spectra
plotted against the temperature of peak line emissivity, taken from
APED.  The open circles represent the Doppler width as measured by
SHERPA.  The filled diamonds represent the rms velocity as measured
by ISIS.  The mean rms velocity and standard deviations of these
lines are indicated by the horizontal lines. Note that two of the lines
formed in the coolest plasma are significantly broader than the
mean, but that most of the lines have non-thermal line widths
of a 250--450~km~s$^{-1}$.
\label{f9}}
\end{figure}

\begin{figure}
\figurenum{10}
\plotone{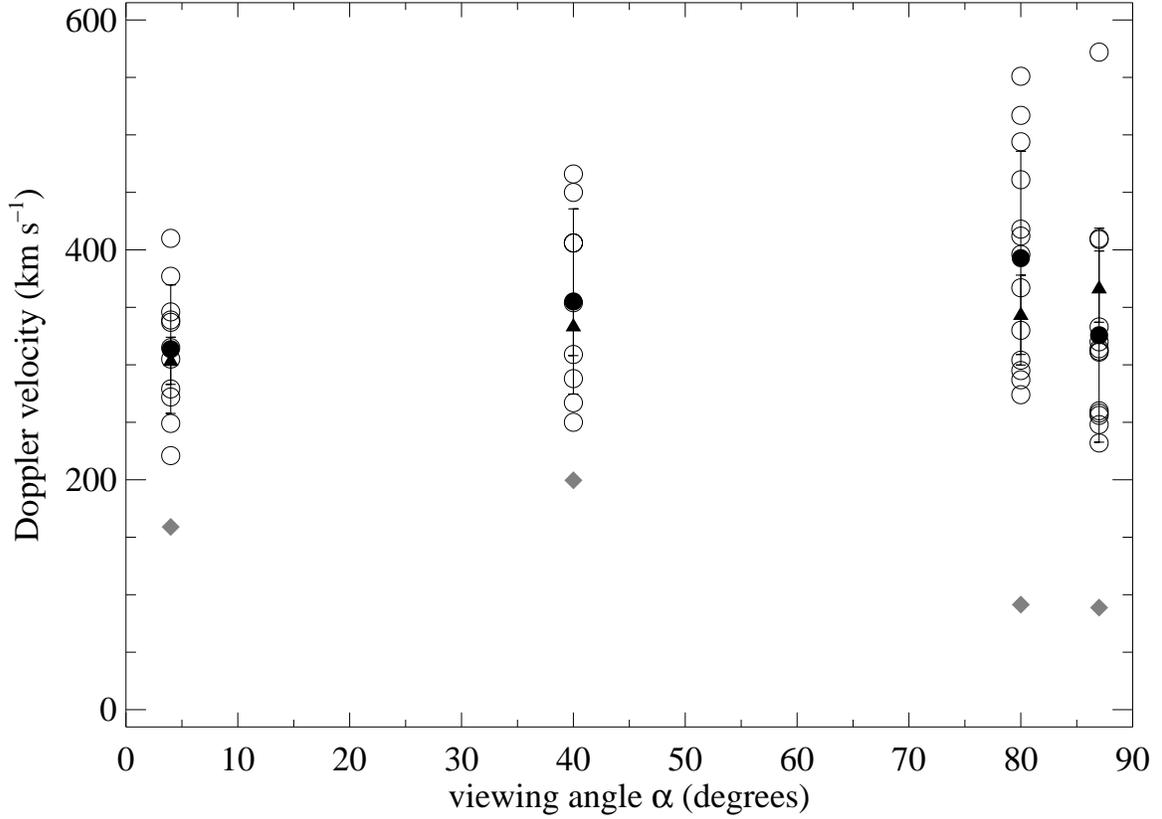}
\caption{
Line widths for the strongest lines in the \chandra\ spectra
(open circles) at each of our four observations.  The filled circles
are the mean measured line widths and the error bars are the standard
deviations of the mean values. The gray diamonds are the theoretical
predictions based on the MHD simulations. The velocity widths plotted
here are the Doppler width components, $v_D$, as described in \S2.
Note that the observed lines are somewhat broader than predicted by
the MHD simulations, but substantially less broad than expected from
line-driven shocks in the wind.
\label{f10}}
\end{figure}

\begin{figure}
\figurenum{11}
\plotone{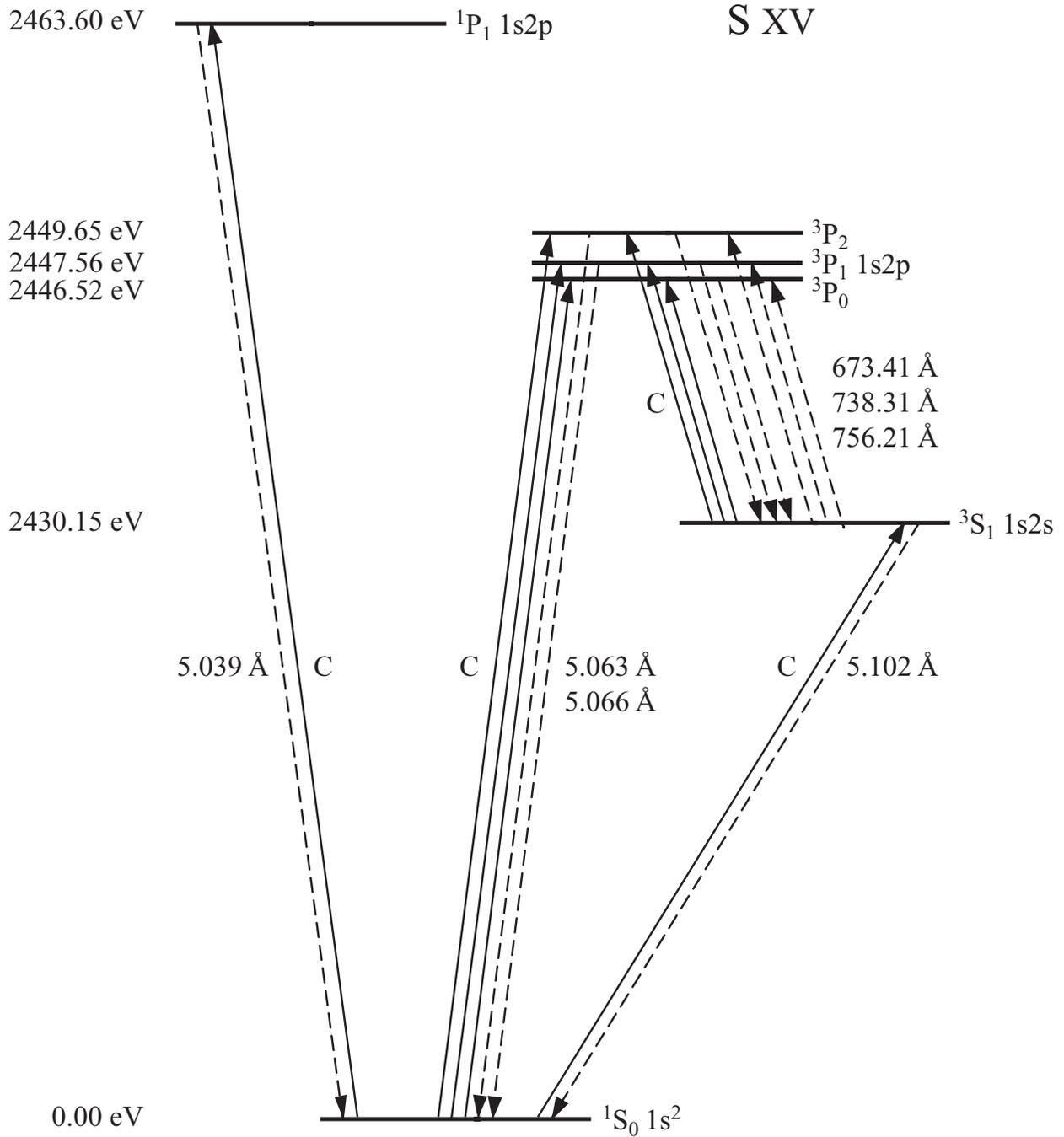}
\caption{
Energy-level diagram for \ion{S}{15},
based on the \ion{O}{7} diagram of \citet{GJ1969}.
Collisional excitations (C) are shown as upward pointing solid lines;
radiative decays are shown as downward pointing dashed lines;
photoexcitations are shown as upward pointing dashed lines.
APED level energies and observed transition wavelengths are also shown.
Electrons are excited out of the metastable $^3{\rm S_1}$ state
by collisions at high density and by radiation close to a hot photosphere.
\label{f11}}
\end{figure}

\begin{figure}
\figurenum{12}
\begin{center}
\includegraphics[scale=0.31]{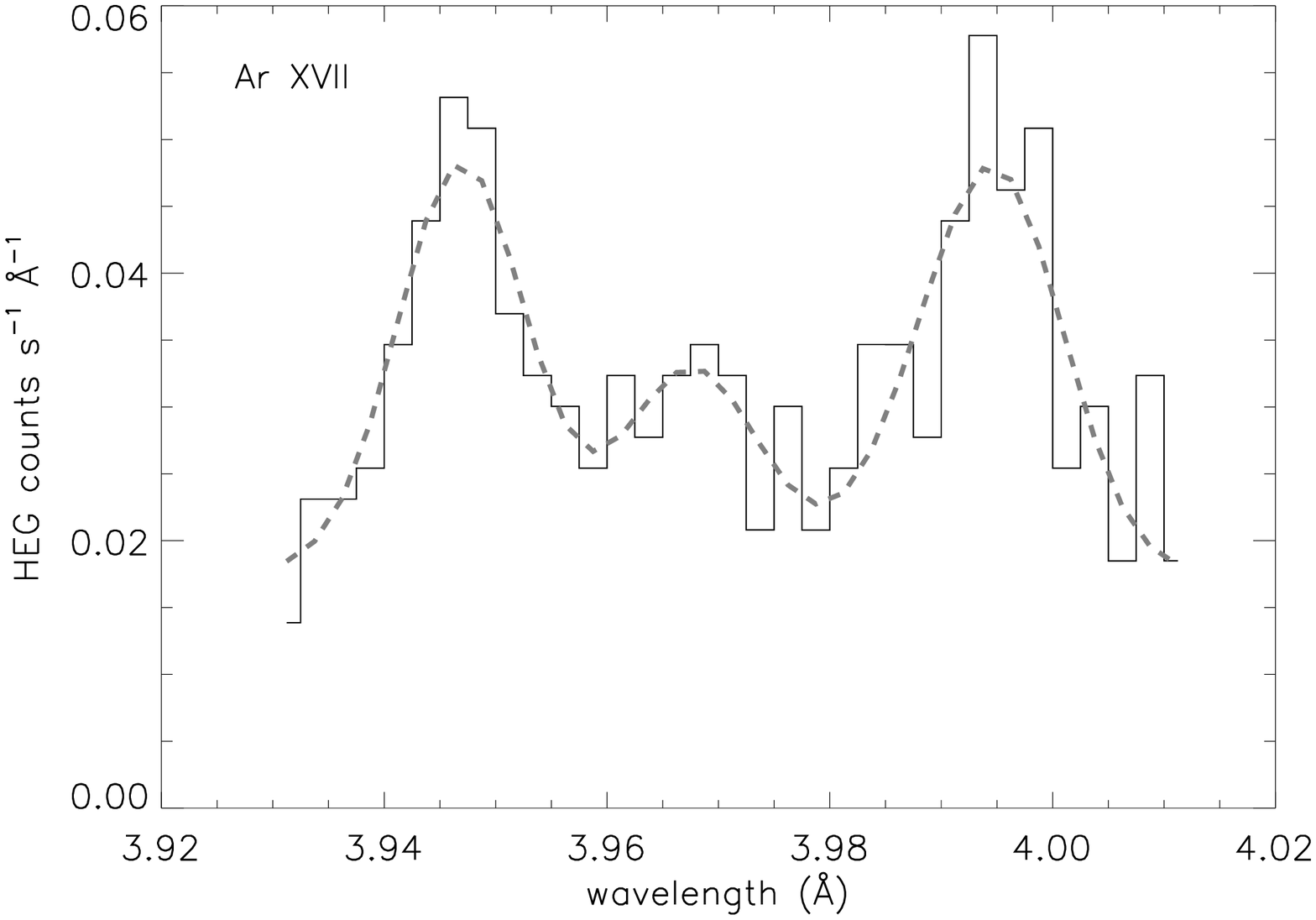}
\includegraphics[scale=0.31]{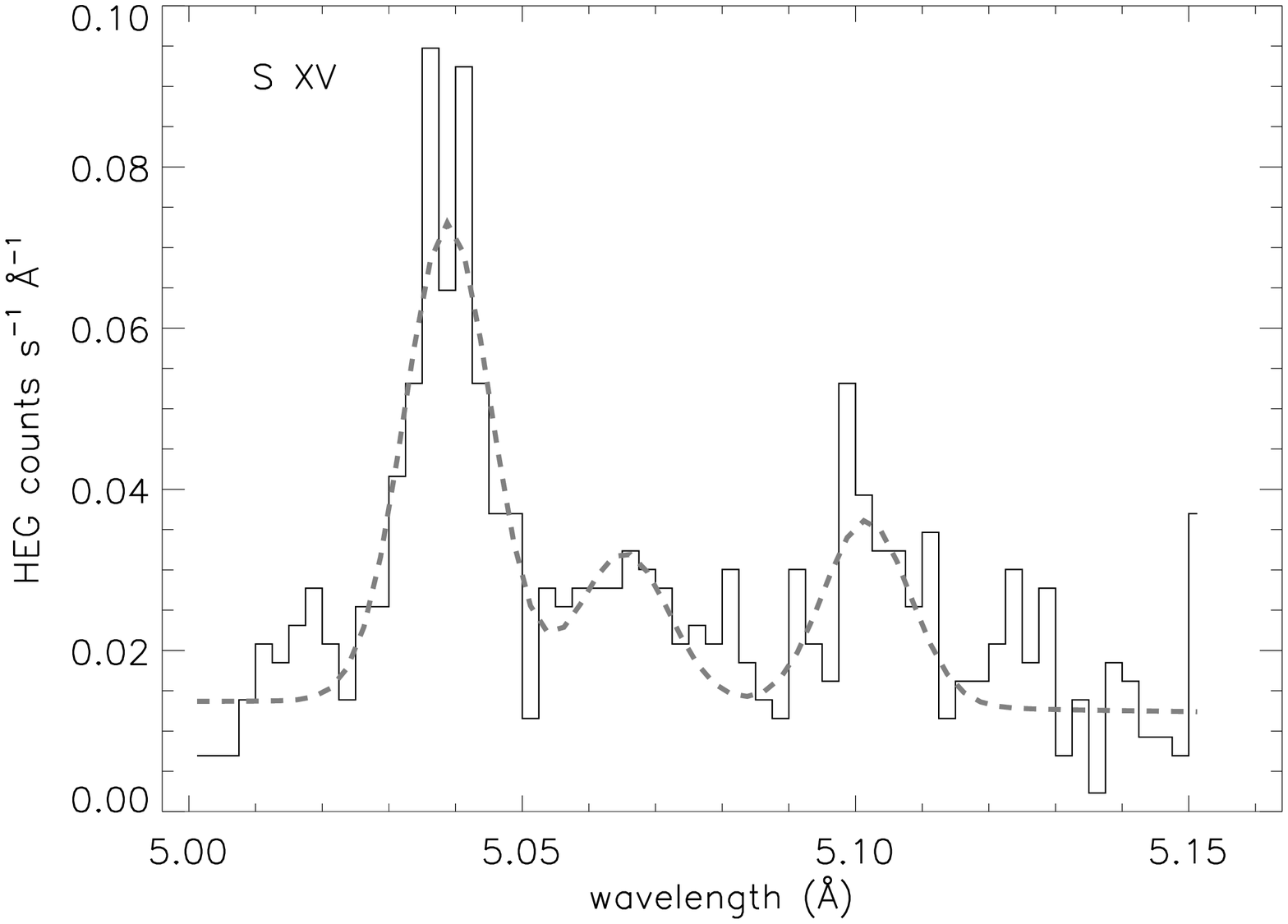}
\includegraphics[scale=0.31]{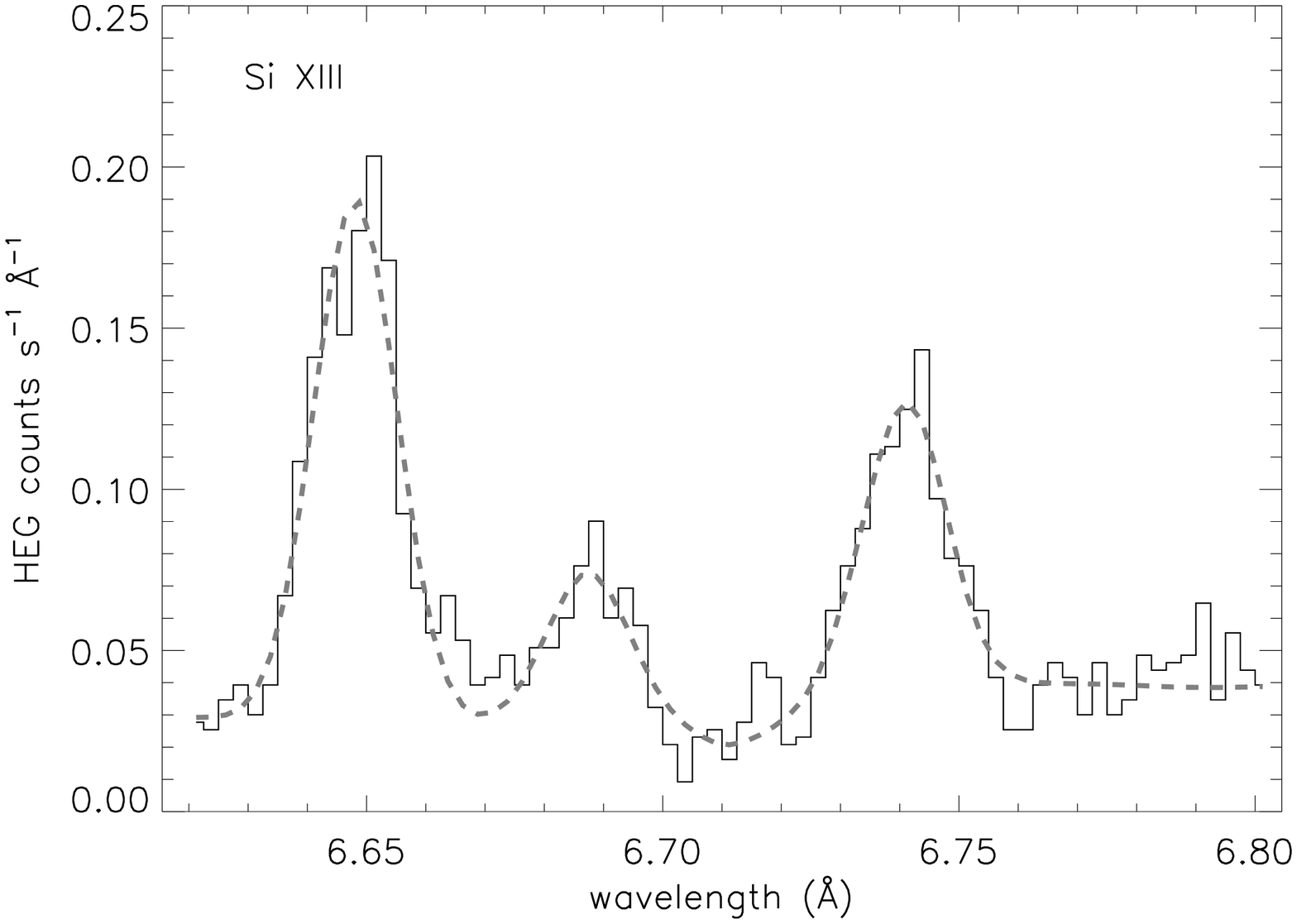}
\includegraphics[scale=0.31]{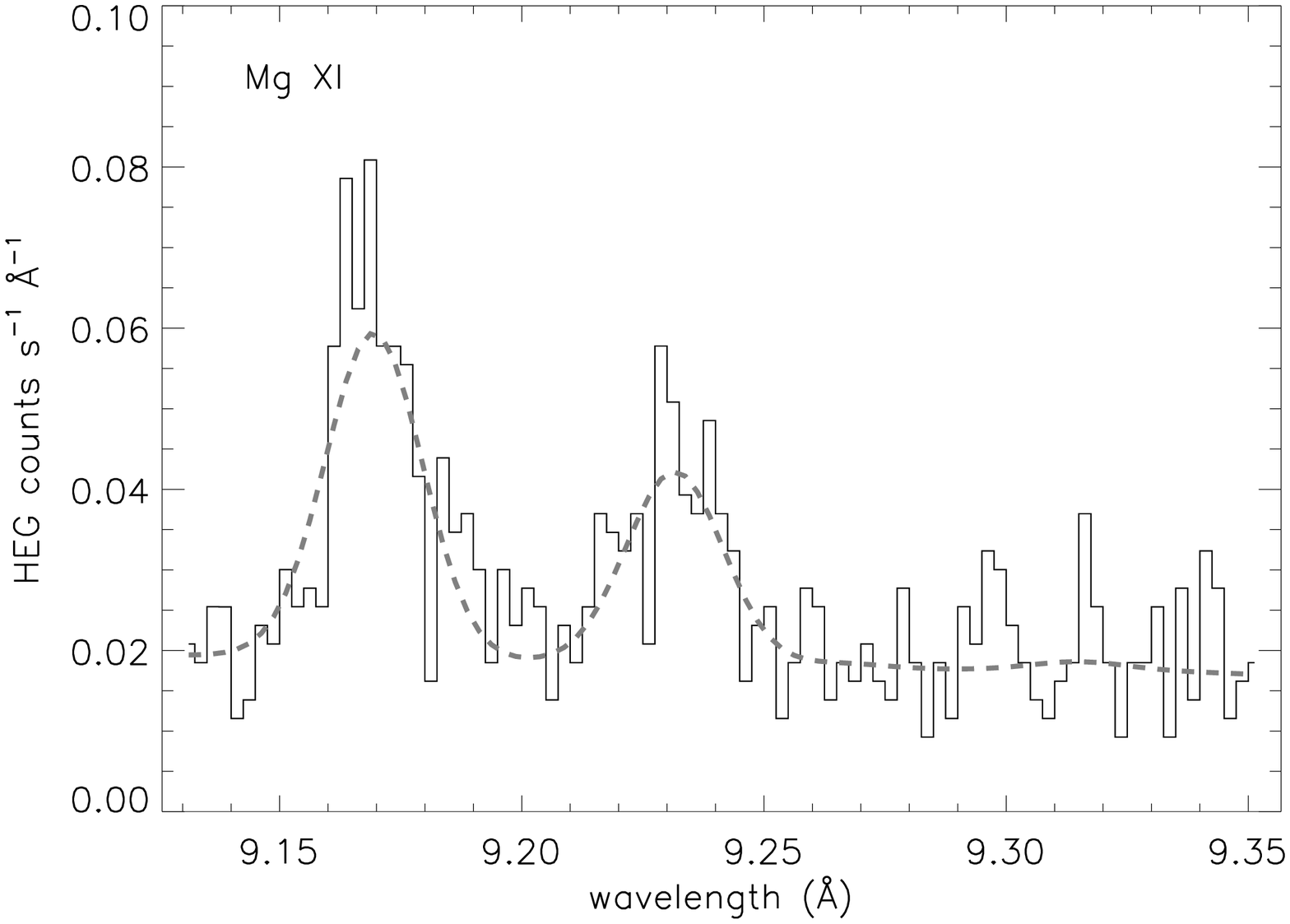}
\end{center}
\caption{
\ion{Ar}{17}, \ion{S}{15}, \ion{Si}{13}, and \ion{Mg}{11}
He-like line complexes.
In each panel, the data (solid histogram) and model (gray dashed line)
of the resonance (r), intercombination (i) and forbidden (f) lines are shown.
Note that as atomic number increases from $Z=12$ (\ion{Mg}{11}) to
$Z=18$ (\ion{Ar}{17}), the relative strength of the
forbidden line to the intercombination line increases.
The \ion{Si}{13} f line is blended with a \ion{Mg}{12} line.
The other detected He-like lines were not used for the diagnostic analysis:
\ion{Ca}{19} was too weak and the \ion{Fe}{25} f and i lines were not resolved.
\label{f12}}
\end{figure}

\begin{figure}
\figurenum{13}
\begin{center}
\includegraphics[scale=0.3]{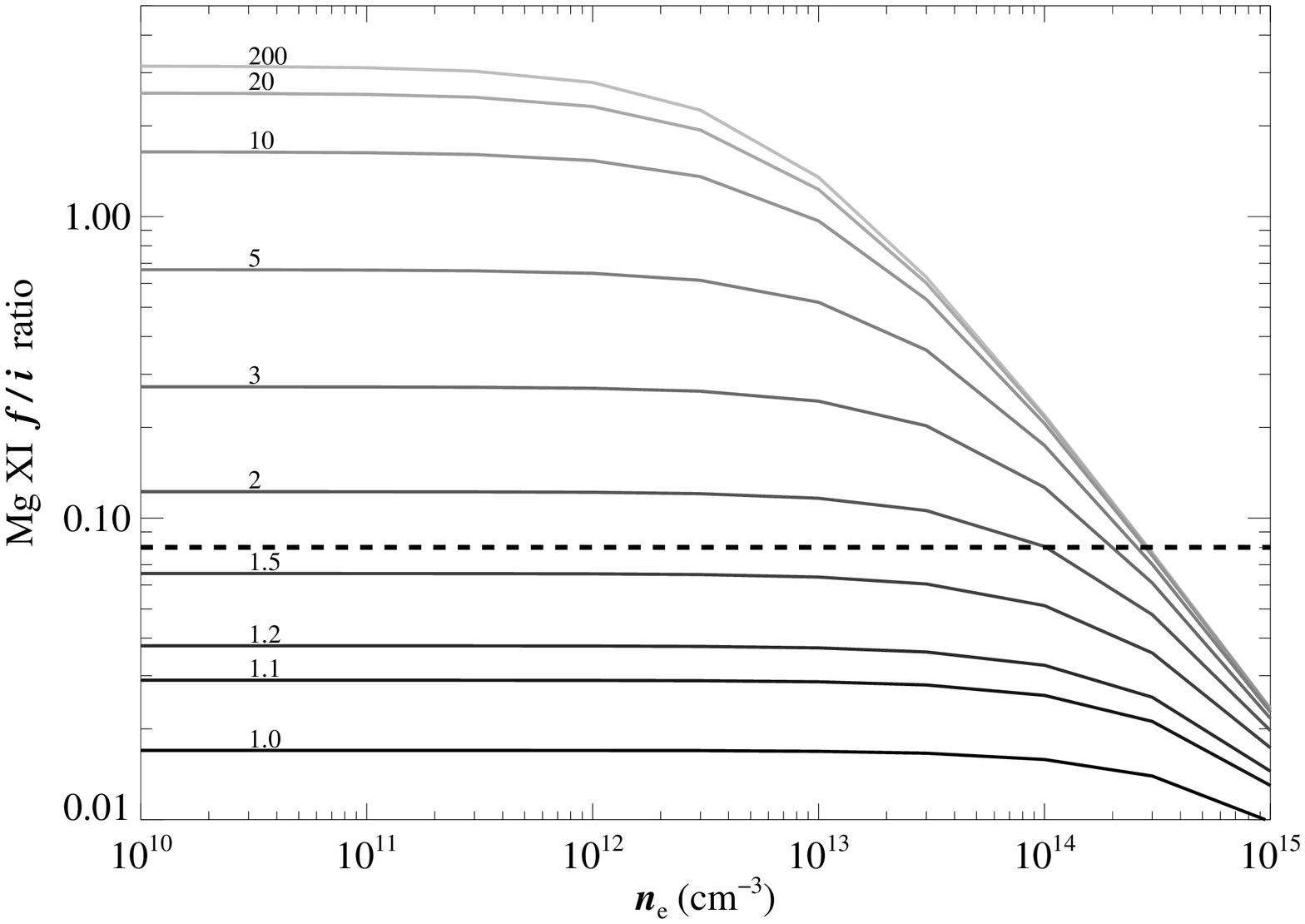}
\includegraphics[scale=0.3]{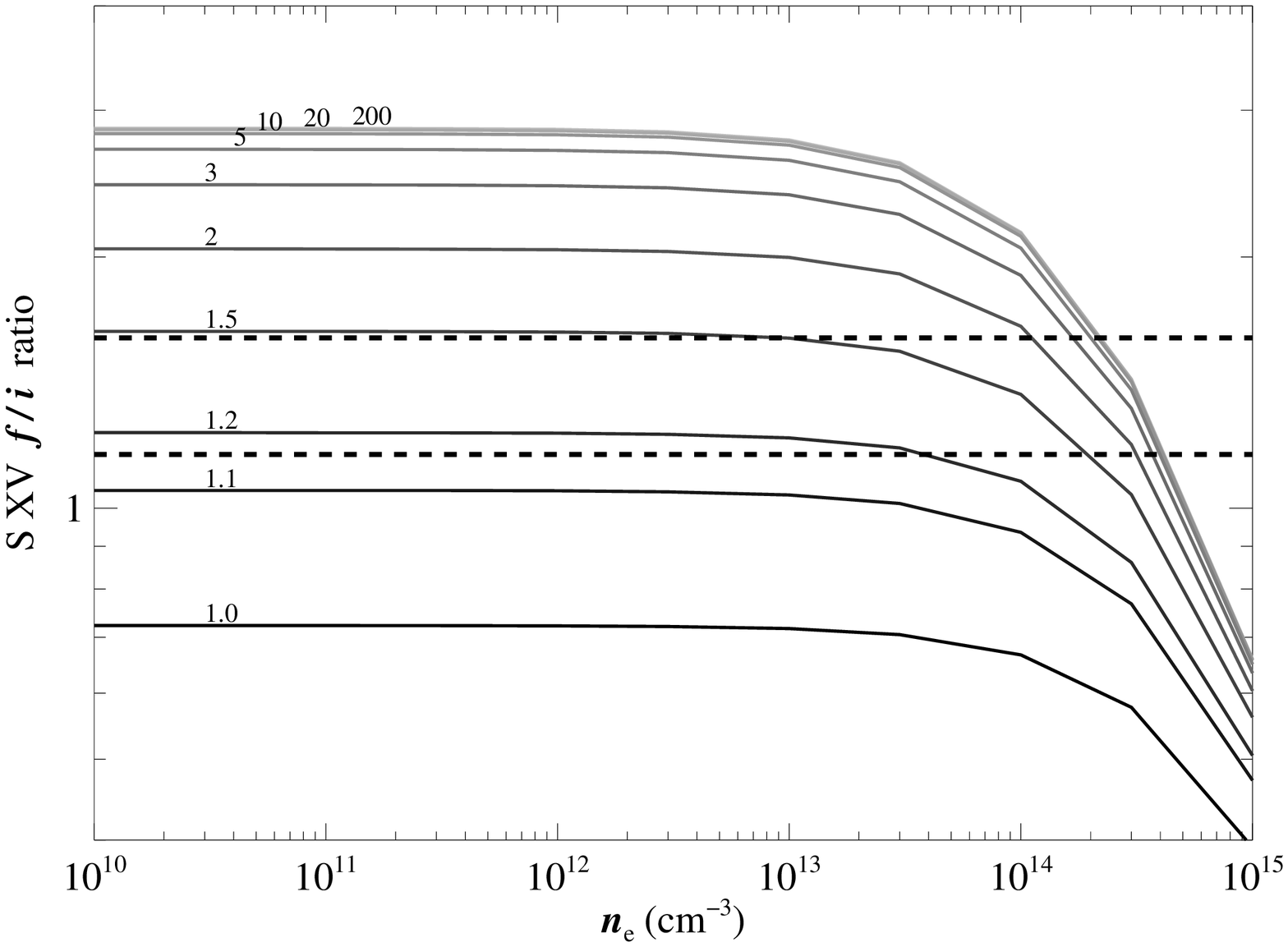}
\end{center}
\caption{
He-like $f/i$ ratio versus $n_{\rm e}$ at various radii from 1--200 $R_\star$,
assuming a 45000~K photosphere, for \ion{Mg}{11} (upper panel)
and \ion{S}{15} (lower panel). The measured $f/i$ upper and lower bounds
from Table~4 are shown as dashed lines. The \ion{Mg}{11}, \ion{Si}{13},
and \ion{S}{15} suggest a formation radius,
$1.2 R_\star \le R \le 1.5 R_\star$.
\label{f13}}
\end{figure}

\begin{figure}
\figurenum{14}
\plotone{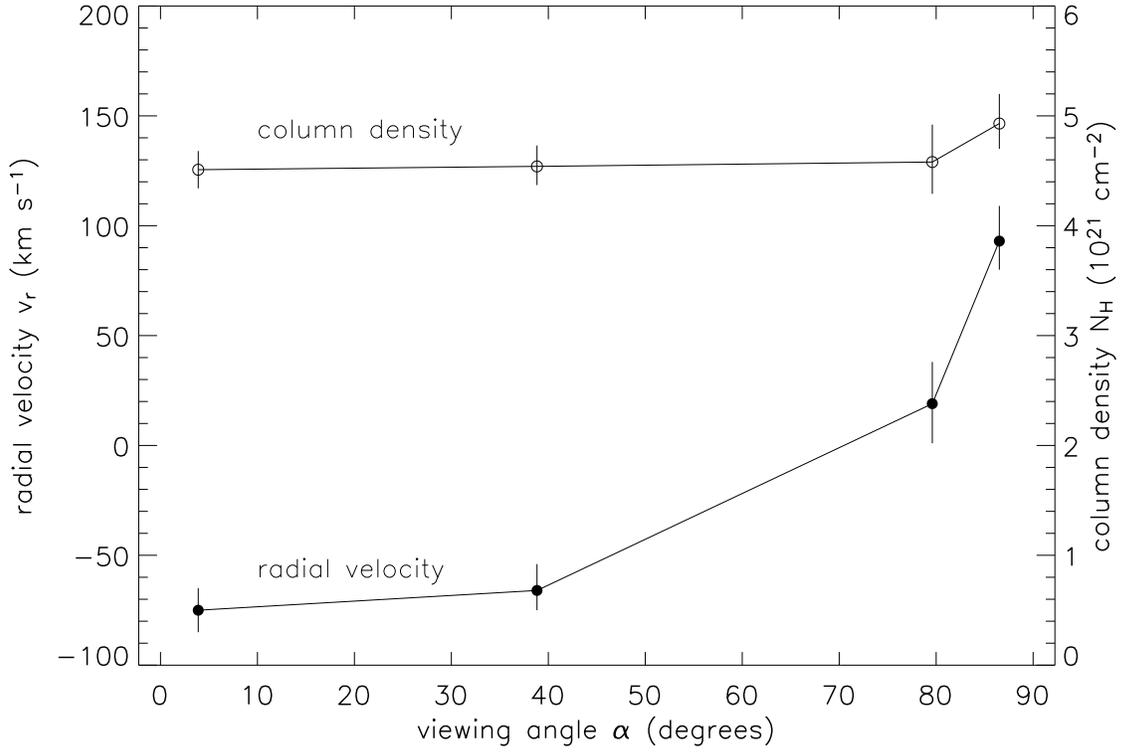}
\caption{
The HEG/MEG spectra show slight changes in radial velocity 
(left axis, filled circles) and column density (right axis, open circles) 
from observation to observation. At low viewing angles (pole-on), the X-ray 
lines are slightly blueshifted; at high viewing angles (disk-on), the lines are 
redshifted with a hint of excess absorption from ionized gas in the disk.
The increased absorption may indicate outflowing wind material in the
magnetic equatorial plane. The redshifts and blueshifts may indicate obliquely
infalling material seen from various viewing angles.
\label{f14}}
\end{figure}

\clearpage
\begin{deluxetable}{lccrrccc}
\tablewidth{0pt}
\tabletypesize{\scriptsize}
\tablecolumns{7}
\small
\tablecaption{{\it Chandra} Observations of $\theta^1$ Ori C}
\tablehead{
\colhead{Sequence}                &
\colhead{Observation}             & 
\colhead{Detector/}               & 
\colhead{Start Date}              &
\colhead{End Date}                &
\colhead{Exposure}                &
\colhead{Average\tablenotemark{1}} &
\colhead{Viewing\tablenotemark{2}} \\ 
\colhead{Number}                  &
\colhead{ID}                      &
\colhead{Grating}                 & 
\colhead{Start Time (UT)}         &
\colhead{End Time (UT)}           &
\colhead{Time (ks)}               &
\colhead{Phase} 		  &
\colhead{Angle}		}   
\startdata
200001  &   3   & ACIS-S HETG &  31 Oct 1999 05:47:21  &   31 Oct 1999 20:26:13  &  \phantom{0}52.0  &  0.84  &  $40^{\circ}$  \\
200002  &   4   & ACIS-S HETG &  24 Nov 1999 05:37:54  &   24 Nov 1999 15:08:39  &  \phantom{0}33.8  &  0.38  &  $80^{\circ}$  \\
200175  & 2567  & ACIS-S HETG &  28 Dec 2001 12:25:56  &   29 Dec 2001 02:00:53  &  \phantom{0}46.4  &  0.01  &  $\phantom{0}4^{\circ}$  \\
200176  & 2568  & ACIS-S HETG &  19 Feb 2002 20:29:42  &   20 Feb 2002 10:01:59  &  \phantom{0}46.3  &  0.47  &  $87^{\circ}$  \\
200214  & 4395  & ACIS-I NONE &  08 Jan 2003 20:58:19  &   10 Jan 2003 01:28:41  &  100.0  & 0.44       &  $85^{\circ}$      \\
200214  & 3744  & ACIS-I NONE &  10 Jan 2003 16:17:39  &   12 Jan 2003 14:51:52  &  164.2  & 0.58       &  $84^{\circ}$      \\
200214  & 4373  & ACIS-I NONE &  13 Jan 2003 07:34:44  &   15 Jan 2003 08:12:49  &  171.5  & 0.75       &  $58^{\circ}$       \\
200214  & 4374  & ACIS-I NONE &  16 Jan 2003 00:00:38  &   17 Jan 2003 23:56:05  &  169.0  & 0.92       &  $19^{\circ}$       \\
200214  & 4396  & ACIS-I NONE &  18 Jan 2003 14:34:49  &   20 Jan 2003 13:15:31  &  164.6  & 0.09       &  $23^{\circ}$       \\
200193  & 3498  & ACIS-I NONE &  21 Jan 2003 06:10:28  &   22 Jan 2003 02:09:20  &  \phantom{0}69.0  &  0.23     & $54^{\circ}$  \\
\enddata
\tablenotetext{1}{Assuming ephemeris $P = 15.422$ d and ${\rm MJD}_0 = 48832.5$ (Stahl et al. 1996).}
\tablenotetext{2}{Assuming inclination $i = 45^{\circ}$ and obliquity $\beta = 42^{\circ}$ (Donati et al. 2002).}
\end{deluxetable}

\begin{deluxetable}{llrrrrrcccc}
\tablewidth{0pt}
\tabletypesize{\scriptsize}
\tablecolumns{11}
\small
\tablecaption{HEG and MEG Line List for $\theta^1$ Ori C: Combined Observations}
\tablehead{
\multicolumn{1}{c}{Ion}              &
\multicolumn{1}{c}{Transition}       &
\multicolumn{1}{c}{$\lambda_0$}      &
\multicolumn{1}{c}{$\lambda$}        &
\multicolumn{1}{c}{$\sigma_\lambda$}        &
\multicolumn{1}{c}{$10^6 f_{\ell}$}  &
\multicolumn{1}{c}{$10^6 \sigma_f$}       &
\multicolumn{1}{c}{$v_{D}$}          &
\multicolumn{1}{c}{$\sigma_{v}$} & 
\multicolumn{1}{c}{$\xi$\tablenotemark{1}}            & 
\multicolumn{1}{c}{$\sigma_{\xi}$}   \\
\multicolumn{1}{c}{}                 &
\multicolumn{1}{c}{}                 &
\multicolumn{1}{c}{(\AA)}                 &
\multicolumn{1}{c}{(\AA)}                 &
\multicolumn{1}{c}{(\AA)}                 &
\multicolumn{2}{c}{(ph~cm$^{-2}$s$^{-1}$)}  &
\multicolumn{2}{c}{(km~s$^{-1}$)} &
\multicolumn{2}{c}{(km~s$^{-1}$)} }
\startdata
Fe XXV   & $1s2p~^1P_{1} \rightarrow 1s^2~^1S_{0}$ & 1.8504  & 1.8498  & 0.0009  & 26.23 & 4.86     & 409 & \nodata & 300 &  \nodata\\
Fe XXV   & $1s2p~^3P_{2} \rightarrow 1s^2~^1S_{0}$ & 1.8554  & 1.8548  & \nodata & 10.00 & 2.62     & 409 & \nodata & 300 &  \nodata\\
Fe XXV   & $1s2p~^3P_{1} \rightarrow 1s^2~^1S_{0}$ & 1.8595  & 1.8589  & \nodata & 9.57  &  \nodata & 409 & \nodata & 300 &  \nodata\\
Fe XXV   & $1s2s~^3S_{1} \rightarrow 1s^2~^1S_{0}$ & 1.8682  & 1.8676  & \nodata & 28.47 & 4.56     & 409 & \nodata & 300 &  \nodata\\
Ca XIX   & $1s2p~^1P_{1} \rightarrow 1s^2~^1S_{0}$ & 3.1772  & 3.1746  & 0.0013  & 8.70  & 1.62     & 426 &   263   & 300 &  \nodata\\ 
Ca XIX   & $1s2p~^3P_{2} \rightarrow 1s^2~^1S_{0}$ & 3.1891  & 3.1865  & \nodata & 4.32  & 0.88     & 426 & \nodata & 300 &  \nodata\\
Ca XIX   & $1s2p~^3P_{1} \rightarrow 1s^2~^1S_{0}$ & 3.1927  & 3.1901  & \nodata & 3.66  &  \nodata & 426 & \nodata & 300 &  \nodata\\ 
Ca XIX   & $1s2s~^3S_{1} \rightarrow 1s^2~^1S_{0}$ & 3.2110  & 3.2083  & \nodata & 6.93  & 1.52     & 426 & \nodata & 300 &  \nodata\\
Ar XVIII & $2p~^2P_{3/2} \rightarrow 1s~^2S_{1/2}$ & 3.7311  & 3.7308  & 0.0016  & 6.48  & 1.16     & 400 & \nodata & 400 &  \nodata\\
Ar XVIII & $2p~^2P_{1/2} \rightarrow 1s~^2S_{1/2}$ & 3.7365  & 3.7362  & \nodata & 3.10  & \nodata  & 400 & \nodata & 400 &  \nodata\\
Ar XVII  & $1s2p~^1P_{1} \rightarrow 1s^2~^1S_{0}$ & 3.9491  & 3.9489  & 0.0009  & 13.02 & 1.91     & 413 &  132    & 300 &  \nodata\\
Ar XVII  & $1s2p~^3P_{2} \rightarrow 1s^2~^1S_{0}$ & 3.9659  & 3.9657  & \nodata & 3.38  & 0.75 & 413 & \nodata & 300 &  \nodata\\
Ar XVII  & $1s2p~^3P_{1} \rightarrow 1s^2~^1S_{0}$ & 3.9694  & 3.9692  & \nodata & 3.46  & \nodata  & 413 & \nodata & 300 &  \nodata\\
S XVI    & $3p~^2P_{3/2} \rightarrow 1s~^2S_{1/2}$ & 3.9908  & 3.9906  & \nodata & 2.44 & \nodata  & 413 & \nodata & 300 &  \nodata\\
S XVI    & $3p~^2P_{1/2} \rightarrow 1s~^2S_{1/2}$ & 3.9920  & 3.9918  & \nodata & 1.22  & \nodata  & 413 & \nodata & 300 &  \nodata\\
Ar XVII  & $1s2s~^3S_{1} \rightarrow 1s^2~^1S_{0}$ & 3.9942  & 3.9940  & \nodata & 12.37 & 1.95     & 413 & \nodata & 300 &  \nodata\\
S XVI    & $2p~^2P_{3/2} \rightarrow 1s~^2S_{1/2}$ & 4.7274  & 4.7270  & 0.0006  & 41.28 & 2.21     & 539 &    77   & 478 &   82   \\
S XVI    & $2p~^2P_{1/2} \rightarrow 1s~^2S_{1/2}$ & 4.7328  & 4.7324  & \nodata & 19.73 & \nodata  & 539 & \nodata & 478 &  \nodata\\
S XV     & $1s2p~^1P_{1} \rightarrow 1s^2~^1S_{0}$ & 5.0387  & 5.0382  & 0.0005  & 50.53 & 3.15     & 331 &    55   & \tablenotemark{*} & \nodata\\
S XV     & $1s2p~^3P_{2} \rightarrow 1s^2~^1S_{0}$ & 5.0631  & 5.0626  & \nodata & 6.54  & 0.98     & 331 & \nodata & \nodata & \nodata\\
S XV     & $1s2p~^3P_{1} \rightarrow 1s^2~^1S_{0}$ & 5.0665  & 5.0660  & \nodata & 9.39  & \nodata  & 331 & \nodata & \nodata & \nodata\\
S XV     & $1s2s~^3S_{1} \rightarrow 1s^2~^1S_{0}$ & 5.1015  & 5.1010  & \nodata & 21.92 & 2.51     & 331 & \nodata & \nodata & \nodata\\
Si XIV   & $2p~^2P_{3/2} \rightarrow 1s~^2S_{1/2}$ & 6.1804  & 6.1803  & 0.0002  & 75.47 & 1.60     & 355 &    23   & 305 &   28    \\
Si XIV   & $2p~^2P_{1/2} \rightarrow 1s~^2S_{1/2}$ & 6.1858  & 6.1857  & \nodata & 36.15 & \nodata  & 355 & \nodata & 305 &  \nodata\\
Si XIII  & $1s2p~^1P_{1} \rightarrow 1s^2~^1S_{0}$ & 6.6479  & 6.6480  & 0.0003  & 61.40 & 1.88     & 370 &    21   & \tablenotemark{*}  & \nodata\\ 
Si XIII  & $1s2p~^3P_{2} \rightarrow 1s^2~^1S_{0}$ & 6.6850  & 6.6851  & \nodata & 14.04  & 1.03     & 370 & \nodata & \nodata & \nodata\\
Si XIII  & $1s2p~^3P_{1} \rightarrow 1s^2~^1S_{0}$ & 6.6882  & 6.6883  & \nodata & 5.40 & \nodata  & 370 & \nodata & \nodata & \nodata\\
Mg XII   & $4p~^2P_{3/2} \rightarrow 1s~^2S_{1/2}$ & 6.7378  & 6.7379  & \nodata & 7.86  & \nodata  & 370 & \nodata & \nodata & \nodata\\
Mg XII   & $4p~^2P_{1/2} \rightarrow 1s~^2S_{1/2}$ & 6.7382  & 6.7383  & \nodata & 3.93  & \nodata  & 370 & \nodata & \nodata & \nodata\\
Si XIII  & $1s2s~^3S_{1} \rightarrow 1s^2~^1S_{0}$ & 6.7403  & 6.7404  & \nodata & 18.98 & 1.45     & 370 & \nodata & \nodata & \nodata\\
Mg XII   & $3p~^2P_{3/2} \rightarrow 1s~^2S_{1/2}$ & 7.1058  & 7.1039  & 0.0009  & 10.97 & 0.77     & 360 &    54   & 311 &    54  \\  
Mg XII   & $3p~^2P_{1/2} \rightarrow 1s~^2S_{1/2}$ & 7.1069  & 7.1050  & \nodata & 5.25  & \nodata  & 360 & \nodata & 311 & \nodata\\ 
Fe XXIV  & $1s^25p~^2P_{3/2} \rightarrow 1s^22s~^2S_{1/2}$ & 7.1690  & 7.1687  & 0.0010  & 8.49  &  0.74   & 360 & \nodata & 311 & \nodata\\ 
Fe XXIV  & $1s^25p~^2P_{1/2} \rightarrow 1s^22s~^2S_{1/2}$ & 7.1690  & 7.1687  & \nodata & 4.15  & \nodata & 360 & \nodata & 311 & \nodata\\ 
Fe XXIV  & $1s^24p~^2P_{3/2} \rightarrow 1s^22s~^2S_{1/2}$ & 7.9857  & 7.9837  & 0.0009  & 13.36 &  0.81   & 363 &    63   & 336 &    61 \\ 
Fe XXIV  & $1s^24p~^2P_{1/2} \rightarrow 1s^22s~^2S_{1/2}$ & 7.9960  & 7.9940  & \nodata & 6.81  & \nodata & 363 & \nodata & 336 & \nodata\\ 
Mg XII   & $2p~^2P_{3/2} \rightarrow 1s~^2S_{1/2}$ & 8.4192  & 8.4183  & 0.0003  & 47.73 & 1.25     & 364 &    20   & 328 &    21  \\  
Mg XII   & $2p~^2P_{1/2} \rightarrow 1s~^2S_{1/2}$ & 8.4246  & 8.4237  & \nodata & 22.91 & \nodata  & 364 & \nodata & 328 & \nodata\\ 
Mg XI    & $1s2p~^1P_{1} \rightarrow 1s^2~^1S_{0}$ & 9.1687  & 9.1693  & 0.0007  & 28.89 & 1.68     & 415 &    49   & \tablenotemark{*}  & \nodata\\
Mg XI    & $1s2p~^3P_{2} \rightarrow 1s^2~^1S_{0}$ & 9.2282  & 9.2288  & \nodata & 2.17  & 0.19     & 415 & \nodata & \nodata & \nodata\\
Mg XI    & $1s2p~^3P_{1} \rightarrow 1s^2~^1S_{0}$ & 9.2312  & 9.2318  & \nodata & 15.20 & \nodata  & 415 & \nodata & \nodata & \nodata\\
Mg XI    & $1s2s~^3S_{1} \rightarrow 1s^2~^1S_{0}$ & 9.3143  & 9.3149  & \nodata & 1.36  & 1.15     & 415 & \nodata & \nodata & \nodata\\
Fe XXIV  & $1s^23p~^2P_{3/2} \rightarrow 1s^22s~^2S_{1/2}$ & 10.6190 & 10.6205 & 0.0006  & 44.26 &  1.77   & 316 &    26   & 288 &   30   \\
Fe XXIV  & $1s^23p~^2P_{1/2} \rightarrow 1s^22s~^2S_{1/2}$ & 10.6630 & 10.6645 & \nodata & 23.28 & \nodata & 316 & \nodata & 288 & \nodata\\ 
Fe XXIII & $1s^22s3p~^1P_{1} \rightarrow 1s^22s^2~^1S_{0}$ & 10.9810 & 10.9821 & \nodata & 25.83 &  1.90   & 300 & \nodata & 263 & \nodata\\
Fe XXIII & $1s^22s3p~^3P_{1} \rightarrow 1s^22s^2~^1S_{0}$ & 11.0190 & 11.0201 & \nodata & 11.39 & \nodata & 300 & \nodata & 263 & \nodata\\
Fe XXIV  & $1s^23d~^2D_{3/2} \rightarrow 1s^22p~^2P_{1/2}$ & 11.0290 & 11.0301 & 0.0007  & 30.29 &  1.61   & 300 &    27   & 263 &   31   \\
Fe XXIV  & $1s^23d~^2D_{5/2} \rightarrow 1s^22p~^2P_{3/2}$ & 11.1760 & 11.1744 & 0.0006  & 42.87 &  2.13   & 236 &    33   & 213 &   38   \\
Fe XXIV  & $1s^23d~^2D_{3/2} \rightarrow 1s^22p~^2P_{3/2}$ & 11.1870 & 11.1854 & \nodata & 4.67  & \nodata & 236 & \nodata & 213 & \nodata\\ 
Fe XXIII & $1s^22s3d~^1D_{2} \rightarrow 1s^22s2p~^1P_{1}$ & 11.7360 & 11.7413 & 0.0006  & 48.87 &  2.78   & 387 &    37   & 387 &   40   \\
Fe XXII  & $1s^22s^23d~^2D_{3/2} \rightarrow 1s^22s^22p~^2P_{1/2}$ & 11.7700 & 11.7753 & \nodata & 25.58 & 2.40 & 387 & \nodata & 387 & \nodata\\ 
Ne X     & $2p~^2P_{3/2} \rightarrow 1s~^2S_{1/2}$ & 12.1321 & 12.1300 & 0.0007  & 56.30 & 2.17     & 351 &    36   & 488 &   35   \\  
Ne X     & $2p~^2P_{1/2} \rightarrow 1s~^2S_{1/2}$ & 12.1375 & 12.1354 & \nodata & 27.05 & \nodata  & 351 & \nodata & 488 & \nodata\\ 
Fe XXIII & $1s^22s3s~^1S_{0} \rightarrow 1s^22s2p~^1P_{1}$ & 12.1610 & 12.1589 & \nodata & 21.17 & \nodata  & 351 & \nodata & 488 & \nodata\\ 
Ne IX    & $1s2p~^1P_{1} \rightarrow 1s^2~^1S_{0}$ & 13.4473 & 13.4386 & 0.0029  & 33.29 & 4.09     & 623 &    86   & \tablenotemark{*} & \nodata \\
Fe XIX   & $2s^22p^33d~^3S_{1} \rightarrow 2s^22p^4~^3P_{2}$ & 13.4620 & 13.4533 & \nodata & 5.93 & 0.61 & 623 & \nodata & \nodata  & \nodata\\
Fe XIX   & $1s^22s^22p_{1/2}2p_{3/2}^23d_{3/2} \rightarrow 2s^22p^4~^3P_{2}$ & 13.4970 & 13.4883 & \nodata & 10.38 & \nodata & 623 & \nodata & \nodata & \nodata\\
Fe XXI   & $1s^22s2p_{1/2}^23s \rightarrow 1s^22s2p^3~^3D_{1}$ & 13.5070 & 13.4983 & \nodata & 9.14 & \nodata & 623 & \nodata & \nodata & \nodata\\
Fe XIX   & $2s^22p^33d~^3D_{3} \rightarrow 2s^22p^4~^3P_{2}$ & 13.5180 & 13.5093 & \nodata & 22.90 & \nodata & 623 & \nodata & \nodata & \nodata\\
Ne IX    & $1s2p~^3P_{2} \rightarrow 1s^2~^1S_{0}$ & 13.5503 & 13.5416 & \nodata & 0.99  & \nodata  & 623 & \nodata & \nodata & \nodata\\
Ne IX    & $1s2p~^3P_{1} \rightarrow 1s^2~^1S_{0}$ & 13.5531 & 13.5444 & \nodata & 25.38 & 4.10     & 623 & \nodata & \nodata & \nodata\\
Ne IX    & $1s2s~^3S_{1} \rightarrow 1s^2~^1S_{0}$ & 13.6990 & 13.6903 & \nodata & 0.41  & 2.77     & 623 & \nodata & \nodata & \nodata\\
Fe XX    & $1s^22s^22p_{1/2}2p_{3/2}3p_{3/2} \rightarrow 2s2p^4~^4P_{3/2}$ & 14.9978 & 14.9993 & \nodata & 0.10 & 0.33 & 614 & \nodata & 700 & \nodata\\
Fe XIX   & $2s^22p^33s~^1D_{2} \rightarrow 2s^22p^4~^1D_{2}$ & 15.0117 & 15.0132 & \nodata & 1.42 & \nodata & 614 & \nodata & 700 & \nodata\\
Fe XVII  & $2s^22p^53d~^1P_{1} \rightarrow 2s^22p^6~^1S_{0}$ & 15.0140 & 15.0155 & 0.0019 & 82.64 & 5.75 & 614 & 67 & 700 & 78 \\
Fe XX    & $1s^22s^22p_{3/2}^23p_{3/2} \rightarrow 2s2p^4~^2D_{5/2}$ & 15.0164 & 15.0176 & \nodata & 0.26 & \nodata & 614 & \nodata & 700 & \nodata\\
Fe XVIII & $2s^22p^43s~^2P_{3/2} \rightarrow 2s^22p^5~^2P_{3/2}$ & 16.0040 & 16.0074 & \nodata & 13.34 & 2.01 & 325 & \nodata & 456 & \nodata \\
O VIII   & $3p~^2P_{3/2} \rightarrow 1s~^2S_{1/2}$ & 16.0055 & 16.0089 & 0.0032 & 5.78 & 2.57 & 325 & 93 & 456 & 67 \\
O VIII   & $3p~^2P_{1/2} \rightarrow 1s~^2S_{1/2}$ & 16.0067 & 16.0101 & \nodata & 2.77 & \nodata & 325 & \nodata & 456 & \nodata \\
Fe XVIII & $2s^22p^43s~^4P_{5/2} \rightarrow 2s^22p^5~^2P_{3/2}$ & 16.0710 & 16.0662 & 0.0042 & 16.41 & \nodata & 325 & \nodata & 456  & \nodata \\
Fe XIX   & $1s^22s^22p_{1/2}2p_{3/2}^23p_{1/2} \rightarrow 2s2p^5~^3P_{2}$ & 16.1100 & 16.1134 & \nodata & 8.90 & \nodata & 325 & \nodata & 456 & \nodata \\
Fe XVII  & $2s^22p^53s~^1P_{1} \rightarrow 2s^22p^6~^1S_{0}$ & 16.7800 & 16.7781 & 0.0033 & 34.25 & 4.87 & 431 & 85 & 456 & \nodata \\
Fe XIX   & $1s^22s^22p_{1/2}2p_{3/2}^23p_{3/2} \rightarrow 2s2p^5~^1P_{1}$ & 17.0311  & 17.0299 & \nodata & 1.29 & \nodata & 313 & \nodata & 290 & \nodata\\ 
Fe XIX   & $1s^22s^22p_{1/2}^22p_{3/2}3p_{1/2} \rightarrow 2s2p^5~^3P_{1}$ & 17.0396  & 17.0382 & \nodata & 0.33 & \nodata & 313 & \nodata & 290 & \nodata\\
Fe XVII  & $2s^22p^53s~^3P_{1} \rightarrow 2s^22p^6~^1S_{0}$ & 17.0510 & 17.0498 & 0.0029 & 32.82 & 5.16 & 313 &  64 & 290 & 87 \\
Fe XVII  & $2s^22p^53s~^3P_{2} \rightarrow 2s^22p^6~^1S_{0}$ & 17.0960 & 17.0948 & \nodata & 29.00 & 4.97 &  313 & \nodata & 290 & \nodata \\
O VIII   & $2p~^2P_{3/2} \rightarrow 1s~^2S_{1/2}$ & 18.9671 & 18.9677 & 0.0037  & 67.49 & 9.54    & 888\tablenotemark{2} & 240     & 851  & 146    \\
O VIII   & $2p~^2P_{1/2} \rightarrow 1s~^2S_{1/2}$ & 18.9725 & 18.9731 & \nodata & 32.53 & \nodata & 888 & \nodata & 851  & \nodata \\
O VII    & $1s2p~^1P_{1} \rightarrow 1s^2~^1S_{0}$ & 21.6015 & 21.5917 & 0.0113  & 16.98 & 9.19 & 292 & \nodata & \tablenotemark{*} & \nodata \\
O VII    & $1s2p~^3P_{2} \rightarrow 1s^2~^1S_{0}$ & 21.8010 & 21.7912 & \nodata & 5.82 & 3.80 & 292 & \nodata & \nodata & \nodata \\
O VII    & $1s2p~^3P_{1} \rightarrow 1s^2~^1S_{0}$ & 21.8036 & 21.7938 & \nodata & 8.60 & \nodata & 292 & \nodata & \nodata & \nodata \\ 
O VII    & $1s2s~^3S_{1} \rightarrow 1s^2~^1S_{0}$ & 22.0977 & 22.0879 & \nodata & 0.00 & 6.20 & 292 & \nodata & \nodata & \nodata \\
\enddata
\tablecomments{Non-zero errors are reported for free parameters only.} 
\tablenotetext{*}{He-like complexes were not fit in ISIS because the APEC model does not account for suppression of the forbidden lines as a result of FUV photoexcitation.}  
\tablenotetext{1}{$\xi$ is the microturbulant velocity as defined by, e.g., Rybicki \& Lightman (1979). Note that the ISIS plasma model explicitly accounts for thermal broadening. Hence, $\xi < v_{D}$.}
\tablenotetext{2}{O VIII was fit using a 2-gaussian model: a narrow component with Doppler width fixed at $v_{D}$ = 300 km~s$^{-1}$ and a broad component with best-fit $v_{D}$ = 1153 km~s$^{-1}$.  $v_{D}$ reported above is the flux-weighted average of the two Doppler widths.} 
\end{deluxetable}

\begin{deluxetable}{lcc}
\tablewidth{0pt}
\tabletypesize{\scriptsize}
\tablecolumns{3}
\small
\tablecaption{Stellar and Wind Parameters}
\tablehead{ \colhead{Property\tablenotemark{*}} & \colhead{Cool Model} &
\colhead{Hot Model} }
\startdata
Spectral Type & O7 V & O5.5 V \\
$T_{\rm eff}$ (K) & 42 280 & 44 840 \\
$\log L$ ($L_{\rm \odot}$) & 5.4 & 5.4 \\
$R$ ($R_{\rm \odot}$) & 9.1 & 8.3 \\
log~$q_0$ (ph cm$^{-2}$ s$^{-1})$ & 24.4 & 24.6 \\
log~$q_1$ (ph cm$^{-2}$ s$^{-1})$ & 23.6 & 23.9 \\
log~$q_2$ (ph cm$^{-2}$ s$^{-1})$ & 17.4 & 18.5 \\
$\mdot$ (\msunyr) & $5.5 \times 10^{-7}$ &  $1.4 \times 10^{-6}$ \\
\vinf\ (km s$^{-1}$) & 2760 & 2980 \\
\enddata
\tablenotetext{*}{$q_0$, $q_1$, and $q_2$ are the numbers of photons per 
unit area per unit time from the stellar surface $R_\star$ shortward of 
912, 504, and 228 \AA, capable of ionizing H I, He I and He II, respectively. }
\label{tab:properties}
\end{deluxetable}

\begin{deluxetable}{lccccccccc}
\tablewidth{0pt}
\tabletypesize{\scriptsize}
\tablecolumns{8}
\small
\tablecaption{He-like Line Diagnostics}
\tablehead{
\colhead{Ion}               &
\colhead{$^3P_2\hspace{-0.5ex}\rightarrow\hspace{-0.5ex}^3S_1$} &
\colhead{$^3P_1\hspace{-0.5ex}\rightarrow\hspace{-0.5ex}^3S_1$} &
\colhead{$r$}                 &
\colhead{$i$}                 &
\colhead{$f$}                 &
\colhead{$R$}           &
\colhead{$\sigma_R$}	    & 
\colhead{$G$}&
\colhead{$\sigma_G$} \\
\colhead{}                  &
\colhead{(\AA)}               &
\colhead{(\AA)}               &
\multicolumn{3}{c}{($10^{-6}$~ph~cm$^{-2}$~s$^{-1}$)}     &
\colhead{$f/i$}                  &
\colhead{} 		    &
\colhead{$\frac{f+i}{r}$}                  &
\colhead{}  }
\startdata
Ar XVII & \phantom{0}560 & \phantom{0}640 & 13.02   &  \phantom{0}6.84  &  12.37   &  1.81    & 0.37    &  1.47    &  0.36   \\
S XV & \phantom{0}673 & \phantom{0}783 & 50.53   &  15.93  &  21.92   &  1.38    & 0.22    &  0.75    &  0.08   \\
Si XIII & \phantom{0}815 & \phantom{0}865 & 61.40   &  19.44  &  18.98   &  0.98    & 0.30    &  0.63    &  0.07   \\
Mg XI & \phantom{0}977 & 1034 & 28.89	&  17.37  &  \phantom{0}1.36   &  0.08    & 0.07    &  0.65    &  0.07   \\
Ne IX & 1248 & 1273 & 33.29	&  26.37  &  \phantom{0}0.41   &  0.02    & 0.11    &  0.83    &  0.19   \\
O VII & 1624 & 1638 & 16.98	&  14.42  &  \phantom{0}0.00   &  0.00    & 0.20    &  0.85    &  0.71   \\
\enddata
\end{deluxetable}

\begin{deluxetable}{lclccll}
\tablewidth{0pt}
\tabletypesize{\scriptsize}
\tablecolumns{7}
\small
\tablecaption{Average Elemental Abundances for $\theta^1$ Ori C}
\tablehead{
\colhead{}                 &
\multicolumn{3}{c}{Global fit}    & 
\multicolumn{2}{c}{Line fit}      &
\colhead{} \\ 
\cline{2-3} \cline{5-6}		\\
\colhead{Element}			&
\colhead{Abundance}		& 
\colhead{$\sigma$}		&
\colhead{}			&
\colhead{Abundance}		&
\colhead{$\sigma$}		&
\multicolumn{1}{l}{Lines used for fit} 			}
\startdata
O	& 0.76 & 0.04 & & 0.76 & 0.08 &O VIII 18.98   \\
Ne	& 0.69 & 0.02 & & 1.04 & 0.05 &Ne X 12.13 \\
Mg	& 0.81 & 0.02 & & 0.94 & 0.02 &Mg XII 8.42	\\
Si	& 1.18 & 0.02 & & 1.11 & 0.02 &Si XIV 6.18 	\\
S	& 1.40 & 0.05 & & 1.22 & 0.06 &S XVI 4.73 	\\ 
Ar	& 1.28 & 0.13 & & 1.48 & 0.17 &Ar XVII 3.95 \\
Ca	& 1.12 & 0.18 & & 1.72 & 0.19 &Ca XIX 3.18 	\\
Fe	& 0.60 & 0.01 & & 0.62 & 0.06 &Fe XVII 15.01, 16.78; Fe XIX 15.01; \\
        &      &      &	&      &      &Fe XX 15.00, 15.02; Fe XXII 11.77; \\
        &      &      & &      &      &Fe XXIII 10.98, 11.02, 11.74; \\
        &      &      & &      &      &Fe XXIV 10.62, 11.03, 11.18; Fe XXV 1.85 \\
\enddata
\tablecomments{Abundances relative to solar (Anders \& Grevesse 1989).  The combined HEG and MEG first-order spectra were fit using a multi-temperature APED plasma model in ISIS.}
\end{deluxetable}

\begin{deluxetable}{llrrrrr}
\tablewidth{0pt}
\tabletypesize{\scriptsize}
\tablecolumns{7}
\small
\tablecaption{Global ISIS fit parameters and 90$\%$ confidence limits}
\tablehead{
\colhead{}				&
\colhead{}				&
\colhead{OBSID 2567}			&
\colhead{OBSID 3}			&
\colhead{OBSID 4}			&
\colhead{OBSID 2568}			&
\colhead{}				\\
\colhead{Parameter}			&
\colhead{Unit}				&
\colhead{$\phi$=0.01}			&
\colhead{$\phi$=0.84}			&
\colhead{$\phi$=0.38}			&
\colhead{$\phi$=0.47}			&
\colhead{Average}			\\
\colhead{}		&
\colhead{}		&
\colhead{$\alpha$=$4^{\circ}$}		&
\colhead{$\alpha$=$40^{\circ}$}		&
\colhead{$\alpha$=$80^{\circ}$}		&
\colhead{$\alpha$=$87^{\circ}$}		&
\colhead{}				}
\startdata
$L_{\rm X}^{\star}$ & $10^{33}$~ergs~s$^{-1}$ & $1.13\pm0.06$ & $1.00\pm0.05$ & $0.77\pm0.04$ & $0.75\pm0.04$ & $1.03\pm0.05$\\
$N_{\rm H}$ & $10^{21}$~cm$^{-2}$&4.51$^{+0.17}_{-0.17}$ &4.54$^{+0.19}_{-0.17}$ &4.58$^{+0.34}_{-0.29}$  &4.93$^{+0.27}_{-0.23}$  &4.61$^{+0.13}_{-0.11}$\\
EM$_{1}$ & $10^{54}$~cm$^{-3}$	&14.3$^{+1.1}_{-0.9}$  &13.6$^{+0.9}_{-0.9}$  &9.7$^{+1.0}_{-1.2}$  &10.3$^{+0.9}_{-0.9}$  &13.7$^{+0.6}_{-0.5}$\\
T$_{1}$ & $10^{6}$ K		&8.4$^{+0.2}_{-0.1}$  &8.4$^{+0.2}_{-0.1}$  &8.4$^{+0.3}_{-0.2}$  &8.5$^{+0.2}_{-0.2}$  &8.5$^{+0.1}_{-0.1}$\\
EM$_{2}$ & $10^{54}$~cm$^{-3}$	&66.2$^{+1.0}_{-1.0}$  &56.5$^{+0.9}_{-0.8}$  &43.3$^{+1.1}_{-1.1}$  &41.3$^{+0.8}_{-0.9}$  &58.5$^{+0.6}_{-0.6}$\\
T$_{2}$ & $10^{6}$ K		&29.2$^{+0.4}_{-0.5}$  &30.1$^{+0.5}_{-0.5}$  &33.6$^{+1.2}_{-1.2}$  &33.6$^{+1.0}_{-1.0}$  &32.6$^{+0.5}_{-0.5}$\\
$\xi$ & km~s$^{-1}$		&303$^{+21}_{-20}$  &333$^{+21}_{-25}$  &343$^{+35}_{-34}$  &366$^{+33}_{-29}$  &348$^{+16}_{-14}$\\
v$_{\rm r}$ & km~s$^{-1}$	&-75$^{+10}_{-10}$  &-66$^{+12}_{-9}$  &19$^{+19}_{-18}$  &93$^{+16}_{-13}$  &-31$^{+10}_{-5}$\\
\enddata
\tablenotetext{\star}{$L_{\rm X}$ is the unabsorbed 0.5--10 keV (1.25--25 \AA) X-ray luminosity assuming $d=450$~pc.}
\end{deluxetable}

\end{document}